\documentclass[preprint2]{aastex62}

\usepackage{natbib}
\usepackage{xcolor}
 
\newcommand\species[2]{#1 {\sc #2}}

\def\ie{\mbox{i.e.}}
\def\eg{\mbox{e.g.}}

\def\teff{\mbox{$T_{\rm eff}$}}
\def\logg{\mbox{log~{\it g}}}
\def\kmsec{\mbox{km~s$^{\rm -1}$}}

\def\vth{$V_{th}$}

\def\vmic{$V_{mic}$}
\def\vmac{$V_{mac}$}
\def\vrot{$V_{rot} {\rm sin}(i)$}
\def\vmacrot{$V_{macrot}$}
\def\fwhmobs{$FWHM_{meas}$}
\def\fwhmplat{$FWHM_{plat}$}
\def\fwhminst{$FWHM_{inst}$}
\def\fwhmunsat{$FWHM_{unsat}$}
\def\sigunsat{$\sigma_{unsat}$}

\shorttitle{Rotation and Macroturbulence in RR Lyrae and RHB Stars}
\shortauthors{Preston et al.}

\begin{document}

\title{THE AXIAL ROTATION AND VARIABLE MACROTURBULENCE OF RR LYRAE 
       AND RED HORIZONTAL BRANCH STARS}

\author{George W. Preston}
\affiliation{Carnegie Observatories, 813 Santa Barbara Street, Pasadena,
             CA 91101, USA; gwp,ian,shec@obs.carnegiescience.edu}
\author{Christopher Sneden}
\affiliation{Department of Astronomy and McDonald Observatory,
             The University of Texas, Austin, TX 78712, USA;
             chris@verdi.as.utexas.edu}
\author{Merieme Chadid}
\affiliation{University of C{\^ o}te d'Azur, Nice Sophia-Antipolis University, 
             ARTEMIS-UMR 7250, CS 34229, 06304 Nice, France;
             Merieme.Chadid@unice.fr}
\author{Ian B. Thompson}
\affiliation{Carnegie Observatories, 813 Santa Barbara Street, Pasadena,
             CA 91101, USA; gwp,ian,shec@obs.carnegiescience.edu}
\author{Stephen A. Shectman} 
\affiliation{Carnegie Observatories, 813 Santa Barbara Street, Pasadena,
             CA 91101, USA; gwp,ian,shec@obs.carnegiescience.edu}

\begin{abstract}

We have derived relations between full-width-half-maxima and equivalent 
widths of metallic absorption lines in the spectra 
of RR~Lyrae stars to estimate new upper limits on the axial 
equatorial rotational velocities of RR~Lyrae and metal-poor red horizontal 
branch stars (RHB).
We also have derived the variations of RR~Lyrae macroturbulent velocities
during the pulsation cycles. 
In RRab cycles the line widths are dominated by phase-dependent 
convolutions of axial rotation and macroturbulence, which we designate as 
\vmacrot. 
The behavior of \vmacrot\ is remarkably uniform among the RRab stars, but
the behavior of \vmacrot\ among RRc stars varies strongly from star to star.
The RRab stars exhibit an upper limit on \vmacrot\ of 5~$\pm$~1~\kmsec\ 
with weak evidence of an anti-correlation with \teff.  
The RRc minima range from 2 to 12~\kmsec. 
The abrupt decline in large rotations with decreasing \teff\ at the blue 
boundary of the instability strip and the apparently smooth continuous 
variation among the RRab and RHB stars suggests that HB stars gain/lose 
surface angular momentum on time scales short compared to HB lifetimes.  
\vmacrot\ values for our metal-poor RHB stars agree well with those derived 
by Fourier analysis of an independent but less metal-poor sample of 
\cite{carney08}; they conform qualitatively to the expectations 
of \cite{tanner13}.
A general conclusion of our investigation is that surface angular momentum 
as measured by \vrot\ is not a reliable indicator of total stellar 
angular momentum anywhere along the HB.
\end{abstract}


\section{INTRODUCTION}\label{intro}

Variations of turbulence during the pulsation cycles of classical 
Cepheids (\citealt{breitfellner93}, \citealt{stift94}, \citealt{bersier96})
and RR Lyrae stars (\citealt{chadid96}, \citealt{fokin99}) have been
well-documented. 
Individual studies have different definitions of turbulence and procedures 
for its extraction from stellar line profiles.
The disentanglement of stellar line broadening due to axial rotation from 
turbulence, however defined, is difficult in all of these investigations, 
which mainly have been concerned with the 
generation of turbulence and its effect on pulsating atmospheres. 

\begin{figure}
\epsscale{1.00}
\plotone{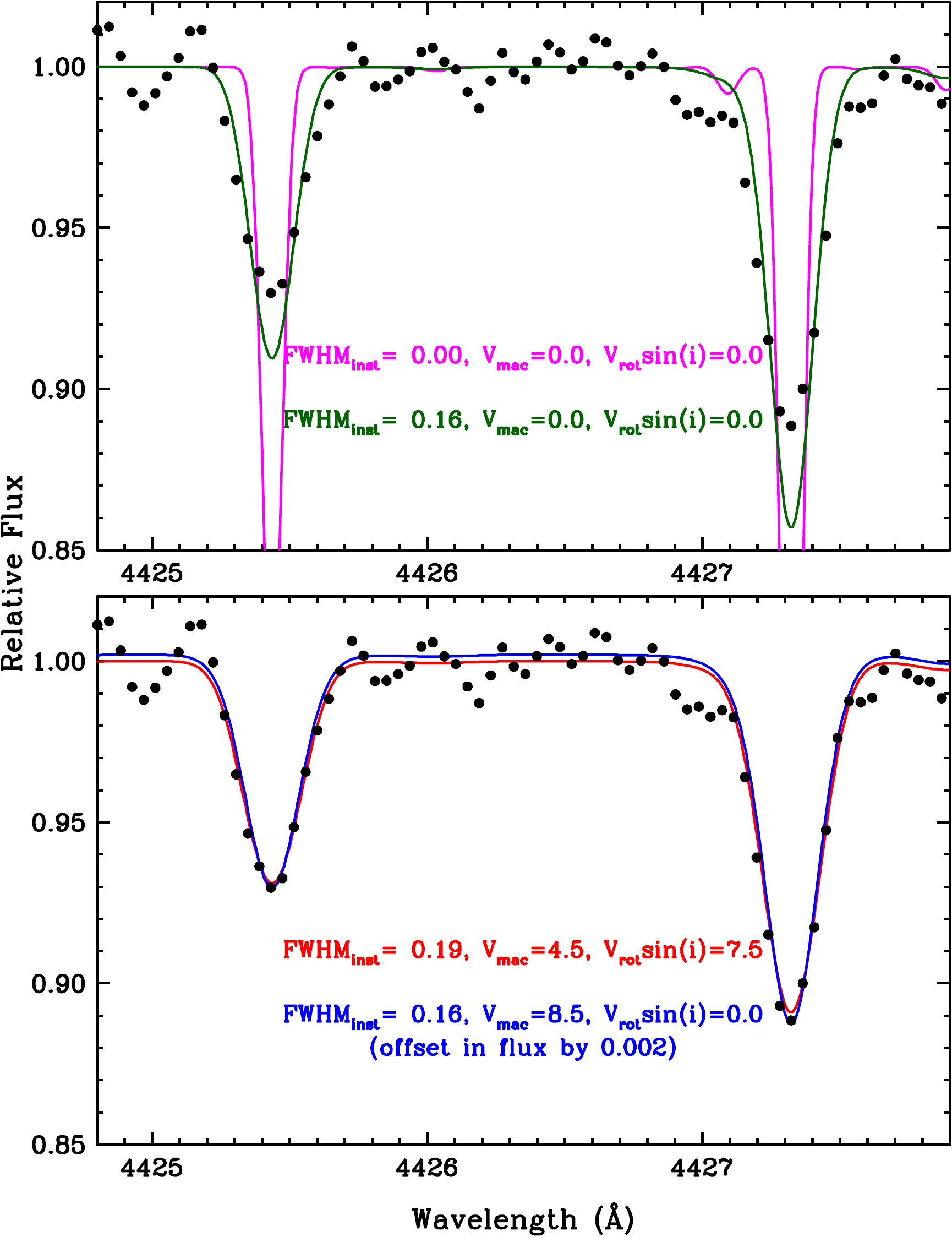}
\caption{\label{fig1}
\footnotesize
   Top panel: Observed and synthetic profiles of \species{Ca}{i} 
   4425.44~\AA\ and \species{Fe}{i} 4427.32~\AA, two unsaturated lines 
   in the spectrum of RRab X~Ari.
   The black dots represent the observed spectrum.
   The magenta line is the raw synthetic spectrum, broadened only by
   thermal and microturbulent velocities assumed in the spectrum
   computations.
   The dark green line shows the raw spectrum broadened by the
   spectrograph slit function only.
   Bottom panel: The same observed spectrum is compared to two synthetic
   spectra. 
   Each of these has the same thermal, microturbulent, and spectrograph
   slit broadening, but the red line has \vmac~=~4.0~\kmsec\ and 
   \vrot~=~6.0~\kmsec, and the blue line has \vmac~=~7.5~\kmsec\ but
   no rotational broadening.
   A small vertical shift has been added to the blue line for plotting clarity.
}
\end{figure}

This problem of disentanglement is illustrated in 
Figure~\ref{fig1}, which shows observed and synthetic spectra of the
X~Ari, a very metal-poor (Fe/H]~=~$-$2.6) RRab star, at phase $\phi$~=~0.37.
In the top panel we show a computed spectrum (magenta color)
of a small spectral region containing the \species{Ca}{i} 4425.44~\AA\ and 
\species{Fe}{i} 4427.32~\AA\ lines.
This spectrum was generated with the synthetic spectrum software and model 
atmosphere parameters described by \cite{chadid17} and transition 
probabilities were adjusted to match the equivalent widths ($EW$s) of the
two main lines.
This computed spectrum only has line broadening that results from thermal
and microturbulent velocity, and clearly does not match the observed
spectrum (black dots).
The dark green spectrum shows this synthesis with added Gaussian smoothing
to account for the du~Pont echelle spectrograph slit function 
($R$~$\equiv$ $\lambda/\Delta\lambda$~$\simeq$ 27,000, or 
full width half maximum $FWHM$~$\simeq$~0.16~\AA\ at 4426~\AA).
This accounts for most but not all of the observed line profiles.
The bottom panel shows two convolutions with identical $FWHM$ values 
but very different Maxwellian macroturbulent and rotational parameters 
(blue, \vmac~=~7.5~\kmsec, \vrot~=~0.0~\kmsec; red, \vmac~=~4.0~\kmsec, 
\vrot~=~6.0~\kmsec).
These profiles appear to be nearly identical in the wings.
Because rotation remains virtually constant during a pulsation cycle, any 
variation of $FWHM$ with phase must be due to thermal and turbulent motions 
of gas elements within RRab atmospheres. 

For 27 RRab stars \cite{peterson96} found a minimum $FWHM$ value near 
pulsation phase $\phi$~=~0.35, the approximate phase of maximum radius, 
derived from cross correlation of metal lines for radial velocity during the 
stellar pulsation cycles.
They argued that the variation of $FWHM$ during pulsation cycles is due to 
variable turbulence and the minimum value is an unresolvable combination 
of turbulence and axial rotation. 
They derived an upper limit of \vrot~$<$~10~\kmsec\ 
for the RR Lyrae stars, similar to that reported for classical 
Cepheids by \cite{bersier96}. 

On the other hand, many blue horizontal branch (BHB) 
stars with \teff~$<$~11500~K\footnote{
As first noted by \cite{grundahl99} HB stars with \teff~$\lesssim$~11,500~K 
have normal (approximately solar) [X/Fe] values, while HB stars hotter than 
this value display abundance anomalies commonly associated with 
radiative levitation and gravitational settling (\eg, \citealt{michaud83},
\citealt{quievy09}, \citealt{theado12}).
 We adopt 11500~K as the upper temperature boundary for BHB stars to be 
included in our discussion.}
are rapid rotators, some attaining \vrot~$\sim$~40~\kmsec\
(\citealt{peterson95}, \citealt{behr03a,behr03b}, \citealt{kinman00}, 
\citealt{recioblanco04}).
Standard HB theory tells us that if these BHB rapid rotators evolve with 
constant Luminosity and constant angular momentum, then their rotation
rates will go as  \vrot~$\propto$~\teff$^2$.
In this case many of them will pass through the instability strip 
rotating more rapidly than \vrot~$\sim$~10~\kmsec.
None of them do so; hence the ``Peterson Conundrum''.

Here we address this puzzle by the study of additional 
samples of RRab and RRc RR Lyrae variables and red horizontal branch
(RHB) stars.
In \S\ref{obsproc} we describe the spectroscopic 
data sets and our extraction of observed broadening of metallic lines for 
each star in individual pulsation phases.
In \S\ref{RRabstable} and \S\ref{RRcRHBvmacrot}, for RRab and RRc stars,
we derive \vmacrot\ quantities, which are combinaions of macro-turbulent and 
rotational broadening effects. 
We compare the \vmacrot\ values to values of micro-turbulent 
velocities \vmic.
In \S\ref{RHBvmacrot} we derive values of \vmacrot\ for very metal-poor
RHB stars and compare them to values for RHB samples used in
previous studies.
Finally, we consider implications of the complex behavior of \vmacrot\
along the horizontal branch in \S\ref{summary}.

\vspace*{0.2in}
\section{OBSERVATIONS AND PROCEDURES}\label{obsproc}

\subsection{The Observational Database}\label{ref}

The main observational data for this study were provided by several thousand 
echelle spectra of 35 RRab stars and 19 RRc stars acquired with the 
du~Pont 2.5~m telescope of the Las Campanas Observatory in the years 
2006$-$2014.\footnote{
Spectra used in this investigation can be obtained at 
https://zenodo.org/record/2575102}
Typical spectral resolving power is, as introduced above,
$R$~$\equiv$~$\lambda/\Delta\lambda$~$\simeq$~27,000 at $\lambda$5000~\AA. 
Exposure times are limited to small fractions of the stellar pulsation
periods, never exceeding 600~s ($\simeq$0.01$P$) and the typical 
signal-to-noise ratio is $S/N$~$\sim$~15 to 20. 
The $S/N$ was improved by factors of two or more at all phases by 
co-addition of spectra in small phase intervals.
Reduction procedures that resulted in wavelength-calibrated extractions 
of du~Pont echelle spectra are described in detail by \cite{for11a} and 
need not be repeated here. 
Basic data for the program stars are given in Table~\ref{tab1}, where stars 
are listed in groups discussed in \S\ref{RRabstable}$-$\ref{RHBvmacrot} below.
For this paper we split the stellar sample into stars
that are metal-poor (MP, [Fe/H]~$<$~$-$1), and those that
are metal-rich (MR, [Fe/H]~$\geq$~$-$1), consistent with the definition
adopted by \cite{chadid17}, \cite{sneden17,sneden18}.
The mean temperatures for the RRab stars in this table 
are taken from \cite{skarka14}, and those for the RRc stars (which have
only small \teff\ variations) are an assumed constant value of 7100~K.
Standard deviations accompany average values when more than one estimate is 
available.

We also obtained many du~Pont echelle spectra of the well known very 
metal-poor ([Fe/H]~$<$~$-$2)
subgiant HD~140283 which we used as a radial velocity standard.
This star has been analyzed repeatedly with high spectral resolution, and
the mean of entries in the PASTEL database \citep{soubiran16} suggests
\teff~=~5700~K, \logg~=~3.6, and [Fe/H]~=~$-$2.4.  
HD~140283 has much narrower absorption lines than RR~Lyrae and other HB stars.

Observations of 24 very metal-poor RHB field stars
stars were obtained with the Magellan MIKE spectrograph 
\citep{bernstein03} in a mode that delivered $R$~$\simeq$~40,000.
The combination of Magellan's larger aperture and longer exposure times
yielded $S/N$~$>$~100 at 5000~\AA\ in all cases.
Reductions and analyses of the Magellan/MIKE spectra are 
discussed in \cite{preston06}.

\subsection{Procedure}\label{proc}

We first derived estimates of line-of-sight velocity dispersion 
caused by the combination of macroturbulence and axial rotation,
after removal of all other measurable line broadening sources. 
In this section we discuss how we avoid line damping problems and how
we correct for instrumental broadening.
Then in \S\ref{RRabstable} we consider the effects of microturbulent 
and thermal broadening.

Cross correlation procedures developed to study galactic 
and stellar rotation (\citealt{simkin74}, \citealt{tonry79,tonry81})
are not useful for RR Lyrae stars.  
During typical RR Lyrae pulsation cycles temperatures vary by as much as 
1000~K, effective gravities vary by more than an order-of-magnitude, and 
turbulent velocities vary by amounts greater than their axial rotations.  
No stars in the CMDs of old metal-poor populations can serve as suitable 
templates for variable stars with these characteristics. 
Accordingly we developed a variant of the method adopted by 
\cite{hosford09}, described in \S\ref{nodamping} below.

\subsubsection{Elimination of Line Damping Effects}\label{nodamping}

For each spectrum we measured $FWHM$ and $EW$ of metal lines chosen from a 
list of some 200 unblended lines (mostly due to \species{Fe}{i}) in the 
wavelength range $\lambda\lambda$4100$-$5300~\AA\ that we judged to be 
unblended by inspection of the solar line identifications of \cite{moore66}.  
These lines are listed in Table~\ref{tab2}.
Less than half of them proved to be measurable in any particular spectrum.
We expect that these lines are unblended in our spectra of RR~Lyrae and RHB
stars and in the spectrum of HD~140283.                
Measurements of these lines are labeled with subscript ``meas''. 
Then, using two methods described below, we culled from these measurements 
subsets of unsaturated  ``plateau'' lines for each of the observed pulsation 
phases of each RR~Lyrae star. 
The measurements of lines surviving this cut are labeled with 
subscript ``plat''.

\begin{figure}
\epsscale{1.00}
\plotone{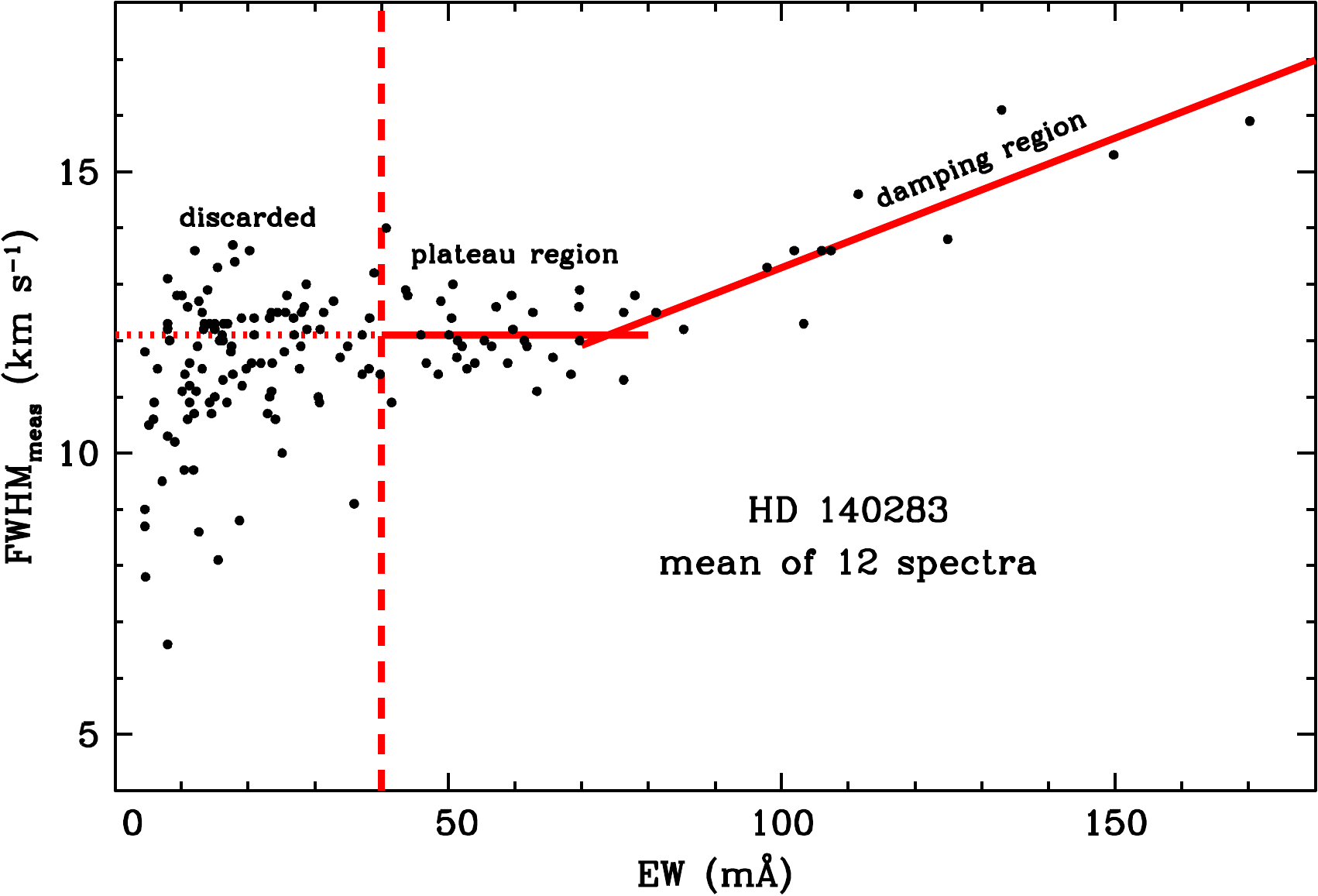}
\caption{\label{fig2}
\footnotesize
   $FWHM_{obs}$ versus $EW$ for lines measured in the du~Pont echelle spectrum 
   of HD~140283.
   These measurements were made on a spectrum formed by merger of 12 separate 
   observations.
   There are three $EW$ regions: the discarded lines, $EW$~$<$~40~m\AA;
   the plateau lines, 40~$\leq$~$EW$~$\leq$~80~m\AA; and damping lines,
   $EW$~$>$~80~m\AA.
   The division of $EW$ regions is discussed in the text.
}
\end{figure}

$(1)$ Our spectra with characteristic $S/N$~$\sim$~20 to 30 and 
resolution of $R$~$\sim$~27,000 are inadequate for direct detection of 
rotation profiles or analysis by Fourier methods.  
Hence, our recourse was to use a variant of the procedure of 
\cite{hosford09}, who determined empirically the average minimum widths of 
weak metal lines as part of their study of lithium in metal-poor main 
sequence turnoff stars (\teff~$\sim$~6200~K, \logg~$\sim$~3.8).
They demonstrated that in a plot of $EW$ versus $FWHM$, lines of their 
stars with $EW$~$<$~90~m\AA\ lie on a ``plateau'' of constant $FWHM$ 
free of detectable damping wings; see their Figure~1. 
We found this procedure to be adequate for our purposes.
We followed \cite{hosford09} in Figure~\ref{fig2} with a plot of 
$FWHM_{obs}$ versus $EW$ for lines of \species{Fe}{i} and \species{Fe}{ii} 
in HD~140283.

In somewhat hotter RR~Lyrae stars, more metal poor by one to two 
orders-of-magnitude, depression of the continuum by hosts of weak unresolved 
lines is of greatly diminished importance. 
The effect of line opacity on the energy distributions of solar-type stars 
was treated in detail long ago by \cite{melbourne60} and \cite{wildey62}.  
Such stars display a marked decline in atomic line blanketing with 
increasing wavelength from 4000~\AA\ to 5500~AA. 
This is illustrated in Figure~1 of \citeauthor{wildey62}, which 
also shows the relative unimportance of line blanketing in the
metal-poor ([Fe/H]~$\sim$~$-$2.0) subdwarf HD~19445.

\begin{figure}
\epsscale{1.00}
\plotone{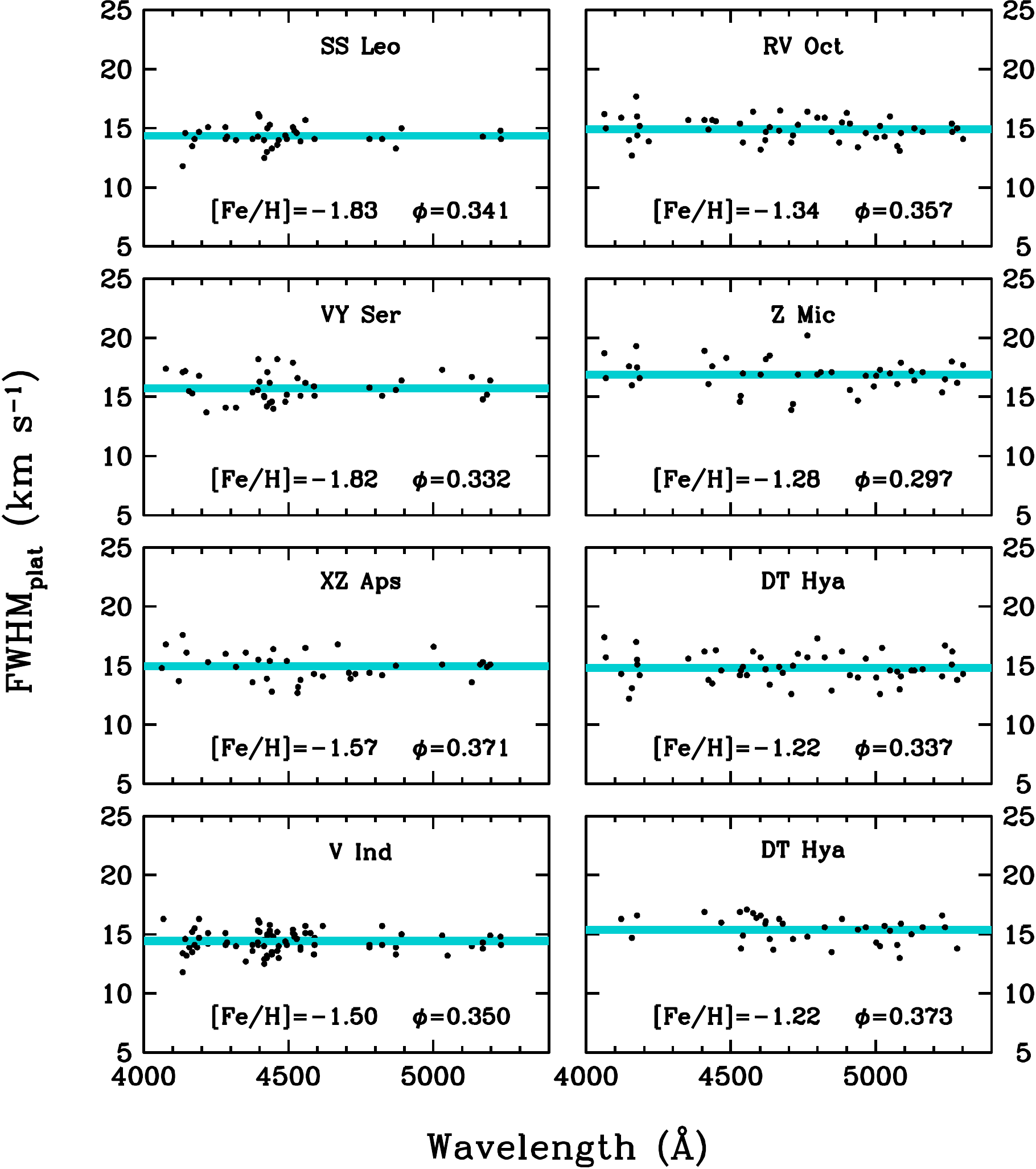}
\caption{\label{fig3}                                         
\footnotesize                                                 
   Plots of FWHM versus wavelength for plateau lines in metal-poor RRab stars, 
   arranged in order of increasing [Fe/H] from top-left to bottom-right.
}                                                             
\end{figure}

All of our measures of $EW$ and $FWHM$ were made by use of the Gaussian 
profile fitting routine of IRAF/splot.  
In this procedure the computer operator chooses the local continuum level at 
each stellar line by interaction with a graphic display of the spectrum.  
If unrecognized metallic lines were preferentially depressing the adopted 
continua at shorter wavelengths, we should see this effect as 
a positive correlation of FWHM with wavelength, but there are no such 
correlations in Figure~\ref{fig3}, displayed for plateau lines at phases 
near \vmacrot\ minimum (phase $\phi$~$\sim$~0.38) in the wavelength range 
4000$-$5300~\AA.  
The stars in this montage are displayed from top-left to bottom-right in 
order of increasing [Fe/H].  
It is instructive to note that the plateau value \fwhmplat~$\sim$~15~\kmsec\
is very nearly the same for all stars plotted in the montage except for Z~Mic, 
a remarkable result that will be further documented in \S\ref{rrabstable}
by the small dispersion of \vmacrot\ minima near phase 0.38.
We conclude that continuum placement is not a problem at phases near \vmacrot\ 
minimum, when the lines are narrowest and of greatest interest to us.

We used an upper bound to the plateau of 80~m\AA\ to identify the onset 
of increases of \fwhmobs\ with increasing $EW$ due to damping in our 
RR~Lyrae spectra. 
Because the \citeauthor{hosford09} spectra had nearly double the
resolving power ($R$~=~47,000) and many times higher signal-to-noise
($\langle S/N\rangle$~$\sim$~100) than those of our RR~Lyrae spectra,
they were able to measure metal-lines as weak as 5~m\AA.
We did not trust our lines with $EW$~$<$~40~m\AA\ to yield reliable $FWHM$
because of a ``personal equation'' effect that we found in our 
measurements of weak lines, \ie, those for which central depths are
comparable to continuum noise of the spectra. 
Accidental negative noise in the wings of a line tends to produce broad 
absorption features that we frequently regarded as unmeasurable. 
However, accidental positive noise in the wings tends to produce 
spuriously narrow but easily measurable features. 
These effects become more important with decreasing spectral resolution 
and $S/N$, becoming particularly noticeable near minimum 
light (pulsation phase $\phi$~$\sim$~0.85), when the lines become broad.  
Decisions about measurability will vary from one person to another, and 
for the same person at different times, hence ``personal equation''. 
The systematic effect arising from these biases, derived from measurements 
of hundreds of lines in eight RRab stars at all pulsation phases, is small 
but detectable (1.07~\kmsec~$\pm$~0.65~\kmsec). 
The effect varies with pulsation phase, going through a minimum near 
phase $\phi$~$\simeq$~0.35 when the lines are intrinsically narrowest.
To summarize, we calculated the average \fwhmplat\ in the more restrictive 
domain 40~m\AA~$<$~$EW$~$<$~80~m\AA. 

\begin{figure}
\epsscale{1.00}
\plotone{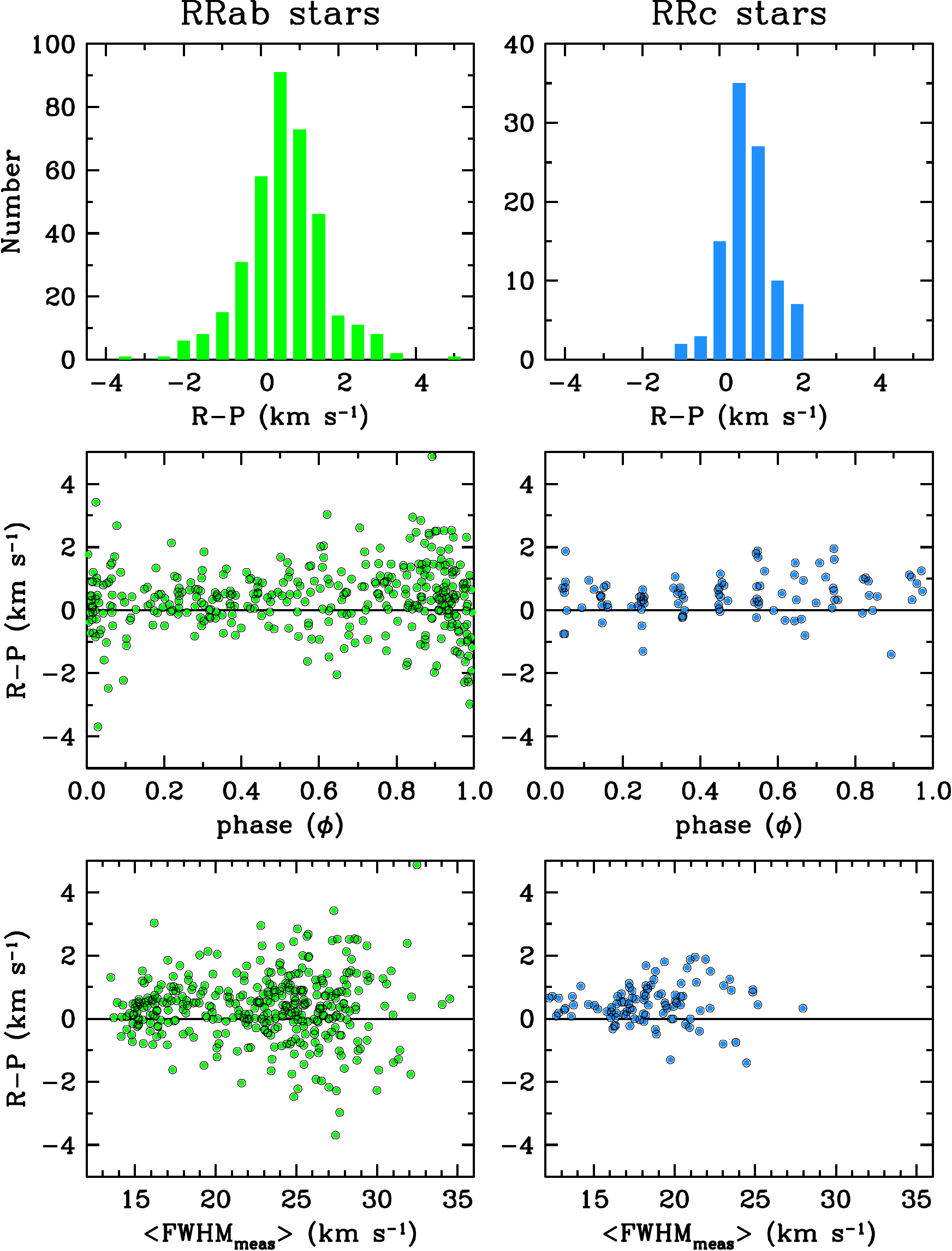}
\caption{\label{fig4}
\footnotesize
   (top) Histogram of the difference $FWHM$ (damping regression at 
   $EW$~=80~m\AA) minus $FWHM$ (plateau) is strongly peaked near 0.5~\kmsec\
   and (middle, bottom panels) varies little with pulsation phase and $FWHM$.
}
\end{figure}

$(2)$ The damping regression for $EW$~$>$~80~m\AA\ produces a second, 
independent estimate of \fwhmplat.
This regression considers stronger, more accurately measured lines than 
those that lie on the plateau. 
Its numerical value, derived by the method of least squares for each spectrum,
depends only slightly on the choice of the plateau $EW$ upper limit. 
The utility of this regression evaluated at $EW$~=~80~m\AA\ as an 
estimator of the plateau value is simply an empirical fact, supported by 
the statistical comparisons of Regression and Plateau estimates in 
Figure~\ref{fig4}.
The top left panel of this figure is a histograms of 370 measured 
differences that we define as
$FWHM({\rm R-P})$~$\equiv$ $FWHM$ (regression at $EW$~=~80~m\AA) $minus$ 
$FWHM({\rm plateau})$ for RRab stars binned at intervals of 0.5~\kmsec.
The average difference, 
$\langle FWHM({\rm R-P})\rangle~=~$0.35~$\pm$~0.05~\kmsec, is 
smaller than the measurement errors.
The lower two panels of Figure~\ref{fig4} show that there is 
little variation of the difference with either pulsation phase or $FWHM$. 
Which of the two estimates is more reliable depends in a complicated way on 
$S/N$ and metallicity. 
The regression values are valuable for observations made during rising 
light and for metal-rich stars, situations in which we frequently are 
unable to measure weak lines.

We adopt the average of these two estimates as our final value of \fwhmplat\
when both are available.

\subsubsection{Correction for Instrumental Line Broadening}\label{instcorr}

Instrumental broadening is comparable to the intrinsic widths of metal 
lines in our RR~Lyrae spectra, so careful removal of it is important for 
the success of our analysis. 
We used the profiles of emission lines produced by a Thorium-Argon hollow 
cathode tube (hereafter ThAr) to remove the effects of instrumental broadening.
The ThAr exposures were obtained by equipment and procedures described below.

The optical paths of starlight and ThAr light for the du Pont 
echelle are nearly identical except for central obscuration of some stellar 
light by the Cassegrain secondary mirror.  
A lens with the focal ratio of the telescope projects an image of the 
ThAr hollow cathode onto the entrance aperture of the spectrograph by a 
motor-operated mirror inserted into the telescope beam.  
ThAr spectra were obtained before and after observations at each position 
of the telescope.  
When we followed a star for lengthy intervals of time, we recorded the ThAr 
spectrum at intervals of 20 to 60 minutes; time 
intervals between ThAr exposures varied as required by various combinations 
of airmass, declination, and hour angle. 
We co-added successive pairs of ThAr exposures to make wavelength-calibrated 
extractions of stellar spectra obtained between successive pairs.

The temperature dependence of the focal adjustment of the echelle 
spectrograph is well-calibrated. 
This calibration was employed to alter the focus of the spectrograph on 
those rare occasions when significant temperature variation during the 
night required its use.

Following the suggestion of an anonymous referee, we also measured the 
$FWHM$ of the [\species{O}{i}] 5577~\AA\ airglow line, which lies some 
200~\AA\ longward of the upper limit of the spectral region used in our 
analyses, on all of the stellar spectra obtained on a dozen nights in 
the years 2006 through 2012.  
We compare these airglow measurements with measurements made of ThAr 
spectra obtained on these same nights in an Appendix.

We approximated corrections for all line broadening processes 
by use of the additive property, 
$\sigma_a$~=~($\sigma_b^2$~$+$~$\sigma_c^2$)$^\frac{1}{2}$, of the Maxwell 
velocity distribution, in which $a,b,c$ are parameters of the Gaussian 
error function. 
We assume that for RR~Lyrae stars the turbulent velocities on all length 
scales are isotropic throughout the metallic line-forming regions.
To begin, we remove the effect of instrumental broadening, characterized by 
\fwhminst, from average values of \fwhmplat\ to obtain 
\fwhmunsat, according to

\begin{center}
\begin{equation}
{\it FWHM_{unsat}} = (\langle{\it FWHM_{plat}}\rangle^2 - \langle{\it FWHM_{inst}}\rangle^2)^\frac{1}{2}
\end{equation}
\end{center}

\noindent in which $\langle$\fwhminst$\rangle$ is an average value described below. 
We then convert \fwhmunsat\ to a Maxwellian velocity dispersion by use of
\sigunsat~=~$(2\sqrt{ln 2})^{-1}$\fwhmunsat.
All further corrections are made in units of line-of sight velocity 
dispersion $\sigma$.

Removal of instrumental broadening was made to the average value 
$\langle$\fwhmplat$\rangle$ rather than to values of individual 
lines, because some measured widths of unsaturated lines may be narrower 
than $\langle$\fwhminst$\rangle$. 
These narrowest lines belong in the average of a normal distribution, but 
they cannot be corrected for instrumental broadening by equation~(1).

\fwhminst\ was evaluated as follows: We measured 591 FWHM values of some 
four dozen lines in each of 14 Thorium-Argon (ThAr) spectra chosen at 
random from spectra gathered in the years 2006$-$2012. 
Details of this investigation are given in the Appendix.  
We used the average value of all these measurements, 
$\langle$\fwhminst$\rangle$ = 11.36~$\pm$~0.52~\kmsec\ in equation (1) to 
calculate \fwhmunsat\ for each RR~Lyrae spectrum. 
This velocity corresponds to resolving power $R$~=~26,700~$\pm$~200, very
close to the value of $R$~$\simeq$~27,000 usually adopted for the du~Pont
echelle spectrograph.

\section{\vmacrot\ FROM MEASUREMENTS OF $EW$ AND \fwhmunsat\ FOR 
         RR{\sc ab} STARS}\label{RRabstable}

For the calibration stars in Table~\ref{tab1} we have enough spectra to 
construct \fwhmobs\ versus $EW$ relations throughout their pulsation cycles.
For each star we combined observations made over several years into 
small phase intervals ($\sim$0.05$P$) to increase $S/N$. 
The average numbers of phase bins for MP and MR stars are 
22 and 18, respectively. 
For each spectrum we used IRAF/splot\footnote{ 
IRAF is distributed by the National Optical Astronomy Observatory, which
is operated by the Association of Universities for Research in Astronomy
(AURA) under cooperative agreement with the National Science Foundation.}
to measure $EW$ and \fwhmobs\ for all usable lines in our list.
Results from these measurements are listed columns 6 through 9 of
Table~\ref{tab1}.

\begin{figure}
\epsscale{1.00}
\plotone{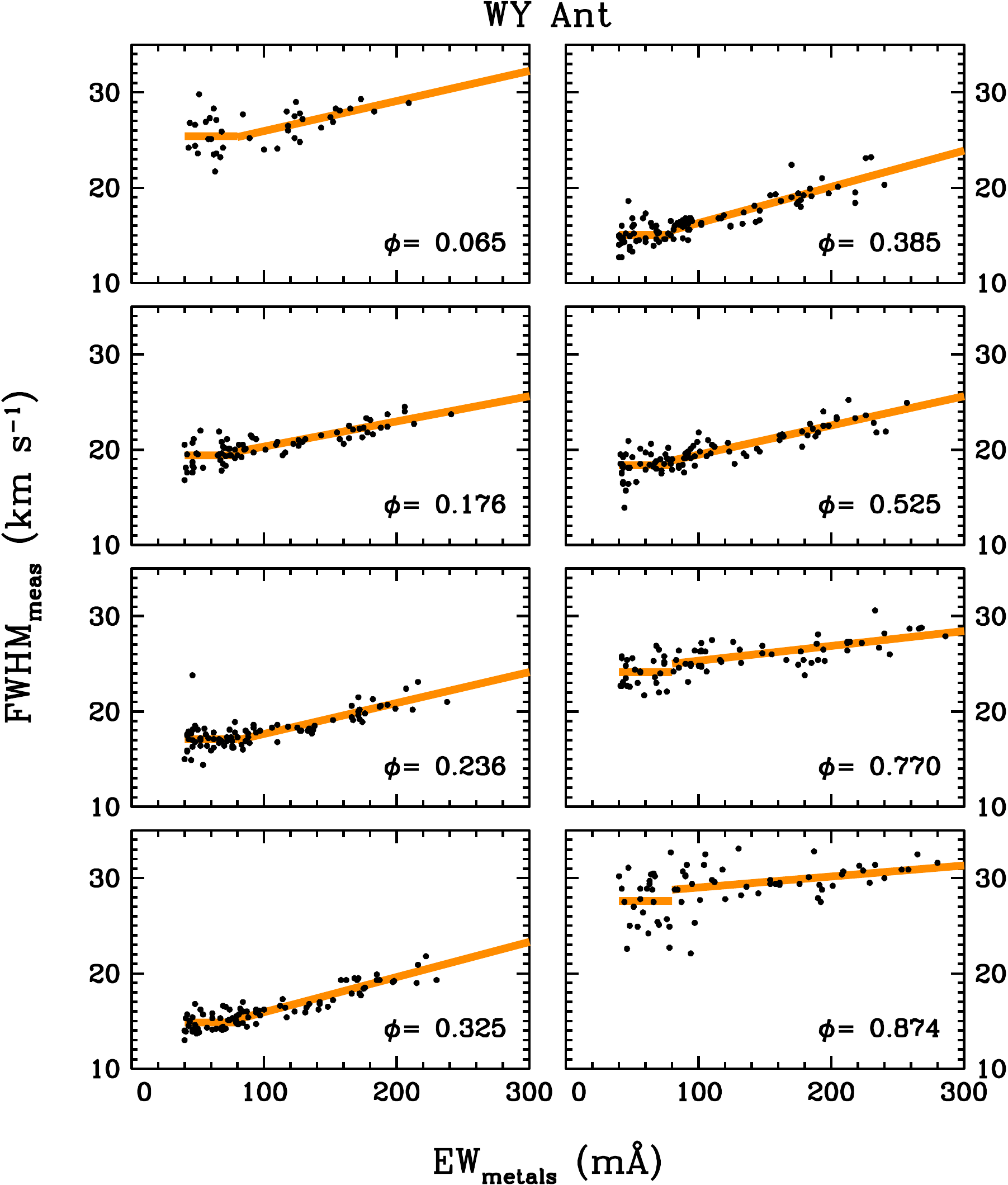}
\caption{\label{fig5}
\footnotesize
   Montage showing how plateau $FWHM_{obs}$ (40~$\leq$~$EW$~$\leq$~80~m\AA) 
   varies during the pulsation cycle of star WY~Ant. 
   Pulsation phases (relative to $\phi$~=~0.0 at maximum light) are shown 
   in the bottom right corner of each panel.
}
\end{figure}

From the \fwhmobs\ versus $EW$ diagrams, direct evidence of 
the variation of macroturbulence with pulsation phase is easily seen. 
In Figure~\ref{fig5} we illustrate this variation for WY~Ant with diagrams 
for eight phases during its pulsation cycle.
The \fwhmobs\ plateau value  is high ($\sim$23~\kmsec) immediately after 
maximum light ($\phi$~=0.065); see the top left panel of Figure~\ref{fig5}. 
It reaches a minimum ($\sim$15~\kmsec) in the bottom left panel, and increases 
steadily during the remainder of declining light shown in the right panels. 

The \fwhmplat\ subsets of these measurements were converted to \fwhmunsat\ 
by equation (1), and from these to \vmacrot. 
The \fwhmplat\ value is similar to the damping regression value at all phases,
as shown in the middle panels of Figure~\ref{fig4}.
In Table~\ref{tab1} we provide estimates of the minimum value of \vmacrot\ 
for each star, calculated as the average \vmacrot\ for observations 
that lie in the phase interval 0.30~$<$~$\phi$~$<$~0.45.

\subsection{The Behaviors of Macroturbulence and Microturbulence in RRab stars}\label{behavior}

\subsubsection{The Stable RRab stars}\label{rrabstable}

Microturbulence, produced by motions on length scales smaller than the 
photon mean free path, broadens all spectral features but for weak
lines (on the linear portion of the COG) the $EW$s are unchanged.
Microturbulence will desaturate lines on the damping portion of the
COG, thus increasing their $EW$s.  
Fortunately, \vmic\ can be derived from spectrum analysis \citep{gray08} 
with standard codes (e.g. MOOG\footnote{
http://www.as.utexas.edu/$\sim$chris/moog.html}
, \citealt{sneden73}).
We used the microturbulence values of \cite{for11b} in our calculations.
Macroturbulence, produced by motions on length scales large compared to 
the photon mean free path, acts to broaden all absorption lines. 
Macroturbulent velocities exceed microturbulent velocities of 
\nocite{richardson20}Richardson's (1920) eddies using the Siedentopf model 
of turbulent convection (\citealt{woolley53}, \citealt{rudiger89}) to which 
we subscribe.

\begin{figure}
\epsscale{1.00}
\plotone{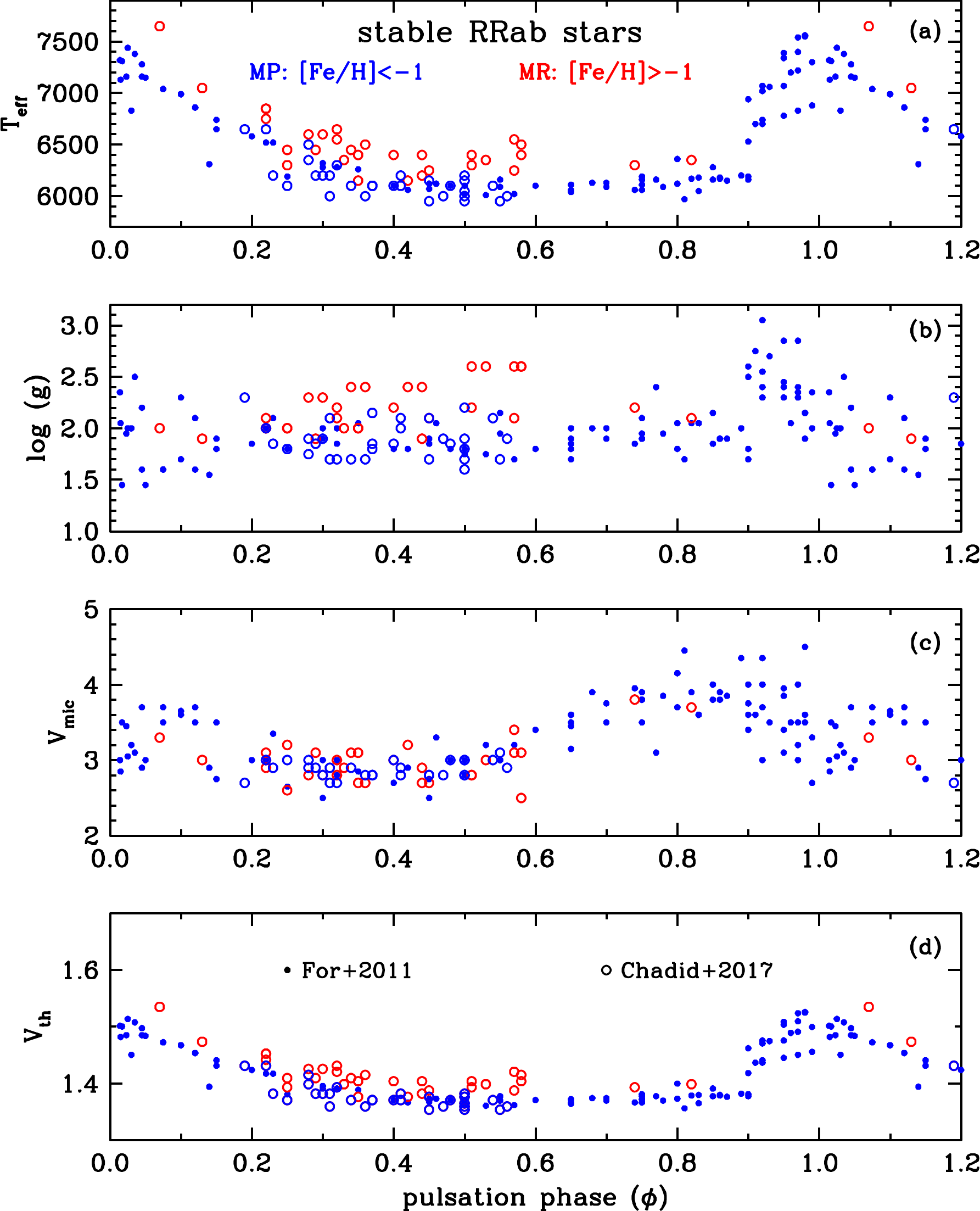}
\caption{\label{fig6}
\footnotesize
   Variations with pulsation phase $\phi$ of \teff\ (panel a), 
   \logg\ (panel b), \vmic\ (panel c), and $v_{th}$ (panel d)
   for the RRab stars studied by \cite{for11b} (dots) and \cite{chadid17}
   (open circles).
   Blue symbols denote MP stars and red symbols denote MR stars.
}
\end{figure}

After computing \sigunsat~=~($2\sqrt{√ln2})^{-1}$\fwhmunsat\ for each 
spectrum of each star, we removed microturbulent and thermal velocity 
dispersions.
These quantities were taken directly or computed from the 
atmospheric parameters for RRab stars derived by \cite{for11b} and 
\cite{chadid17}.  
In Figure~\ref{fig6} we show the variations in \teff, \logg, 
\vmic, and \vth\ as functions of pulsation phase for these stars.
The gravities (panel b) are based on ionization equilibrium constraints
only, as discussed in \citeauthor{for11b} and \citeauthor{chadid17}.
Their values are shown here only for completeness, as we will not use them
in our subsequent calculations.
Additionally, all quantities show significant scatter in the phase domains
of rising and maximum light (0.8~$<$~$\phi$~$<$~1.1); these parts of the
pulsational cycles will not be used here.

\begin{figure}
\epsscale{1.00}
\plotone{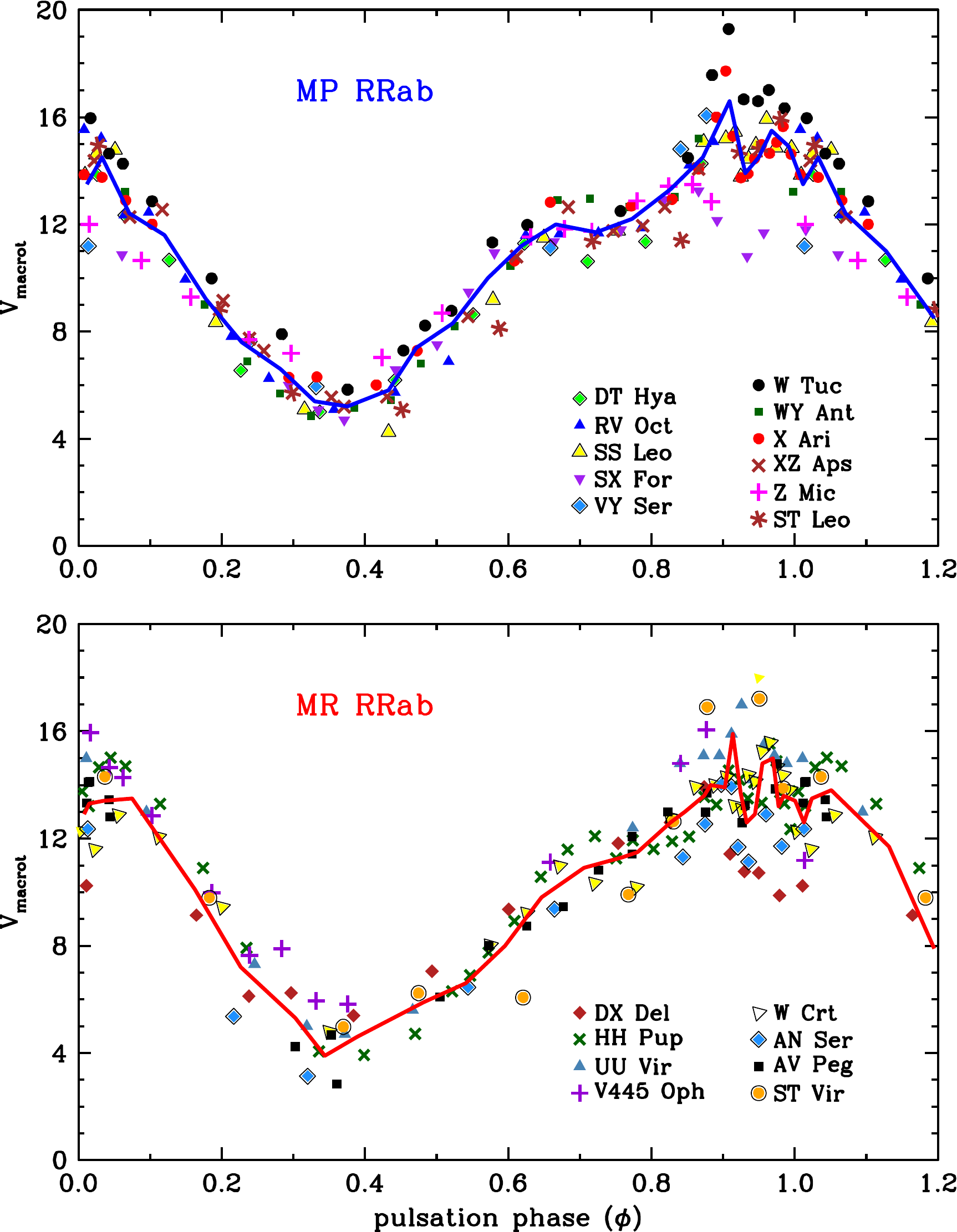}
\caption{\label{fig7}
\footnotesize
   Variations of \vmacrot\ with phase for MP (top panel) 
   and MR (bottom panel) RRab stars.
   Solid curves are mean variations as described in the text.
}
\end{figure}

We computed thermal velocities from \vth~=~(2kT/m)$^\frac{1}{2}$, 
adopting m~=~54~amu\footnote{
54~amu is the mean value of the atomic masses of lines participating in the
velocity calculations.},
and these values are shown in panel (d) of Figure~\ref{fig6}.
These velocities are nearly constant, ranging only over
1.38~$\lesssim$~\vth~$\lesssim$~1.53, due to the modest RRab temperature 
variations (panel a) that are muted by the square root function.
Over most of the pulsational cycles 
\vth~$\simeq$~1.4~\kmsec.
Additionally, the microturbulent velocities shown in panel (c) are 
consistently larger: \vmic~$\simeq$~2\vth\ at all phases, so we have
assumed a simgle microturbulent velocity 
of 2.8~\kmsec\ in all subsequent calculations.
We use these data to plot
\vmacrot~=~(\sigunsat$^2 - \sigma_{mic}^2 - \sigma_{th}^2)^\frac{1}{2}$
versus phase in Figure~\ref{fig7}. 
The top and bottom panels contain, respectively, data for MP and MR stars.

\begin{figure}
\epsscale{1.00}
\plotone{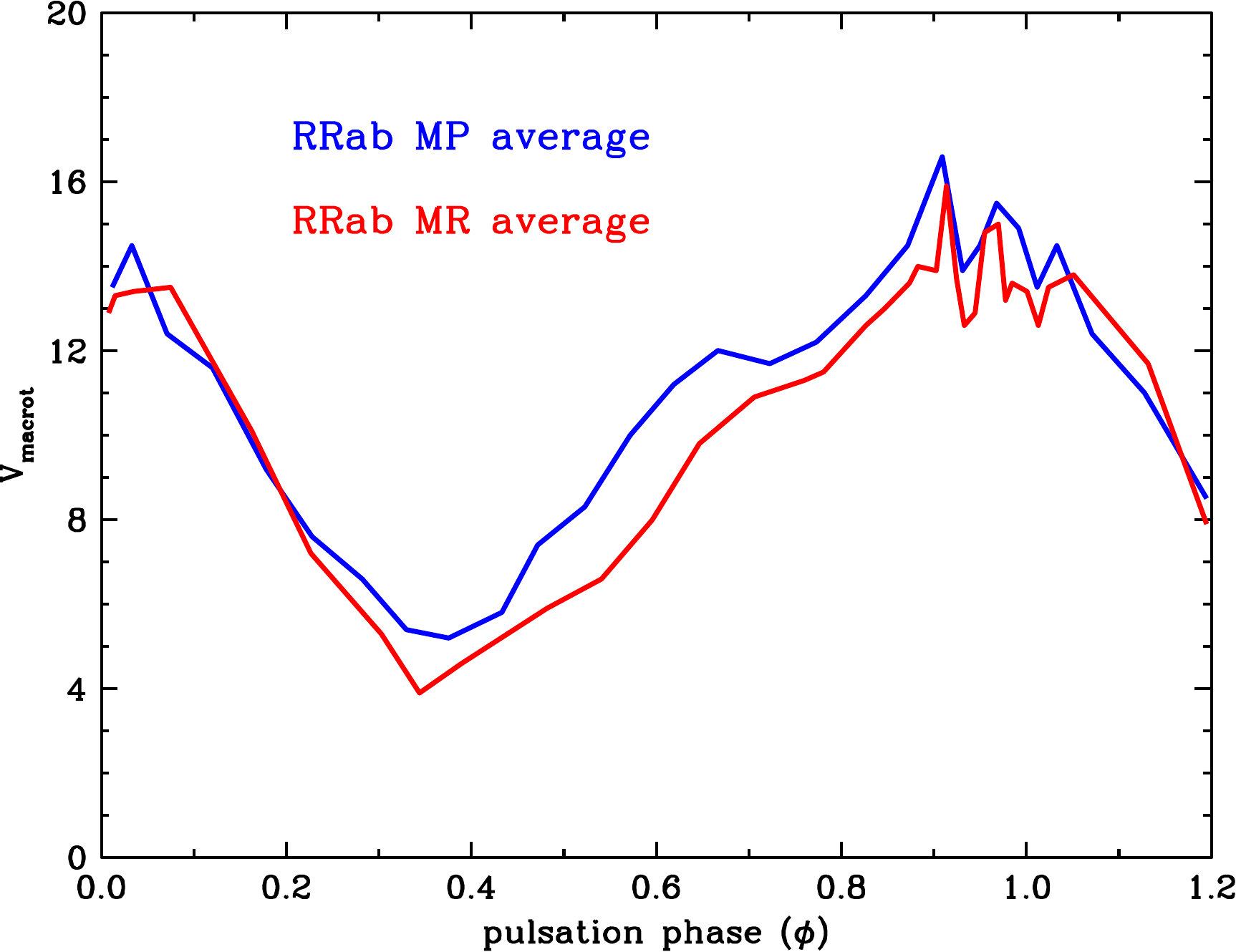}
\caption{\label{fig8}
\footnotesize
   Average variations of \vmacrot\ with phase for MP (blue) and MR (red) 
   RRab stars. 
   The minimum value of \vmacrot/ for MR stars is 
   significantly lower than that for MP stars and the rising 
   branch is shifted to later phase.
}
\end{figure}

We also calculated average variations of \vmacrot\ with phase, the solid 
curves in Figure~\ref{fig7}, as follows. 
We sorted all the data by phase and then calculated averages
$\langle\phi\rangle$ and $\langle$\vmacrot$\rangle$ for data binned 
in successive groups of 10 measurements. 
We superpose the mean MP and MR curves without the individual data points in 
Figure~\ref{fig8}. 
Minimum \vmacrot\ values for all the RR~Lyrae stars in this 
investigation and for the MP RHB stars to be
discussed in \S\ref{RHBvmacrot} are presented in Table~\ref{tab1}.
The minimum averages, computed from data for each RRab in the phase interval
0.30~$<$~$\phi$~$<$~0.45, are 5.31~$\pm$~0.13~\kmsec\ and
4.40~$\pm$~0.22~\kmsec, respectively.
These averages differ by an amount three times larger than the sum of 
their mean probable errors. 
Furthermore, the persistent offset of the two variations after phase 
$\phi$~=~0.5 in Figure~\ref{fig8} indicates a small but real 
difference between the behaviors of \vmacrot\ in MP and MR stars. 

Although we can only claim that the \vmacrot\ minima in Figure~\ref{fig7} 
provide upper limits on \vrot\ for the MP and MR RRab, the small scatter of 
individual stars near these minima strongly suggests that macroturbulence, 
not rotation, dominates the minimum values.  
Random inclination of rotation axes alone should create dispersion in the
minimum values, and we are uncomfortable with the notion that all RRab stars 
rotate at the same speed: thus, the actual rotations of RRab stars may 
lie well below our calculated upper limits.

\subsubsection{The Blazhko RRab stars}\label{rrabblazhko}

\begin{figure}
\epsscale{1.00}
\plotone{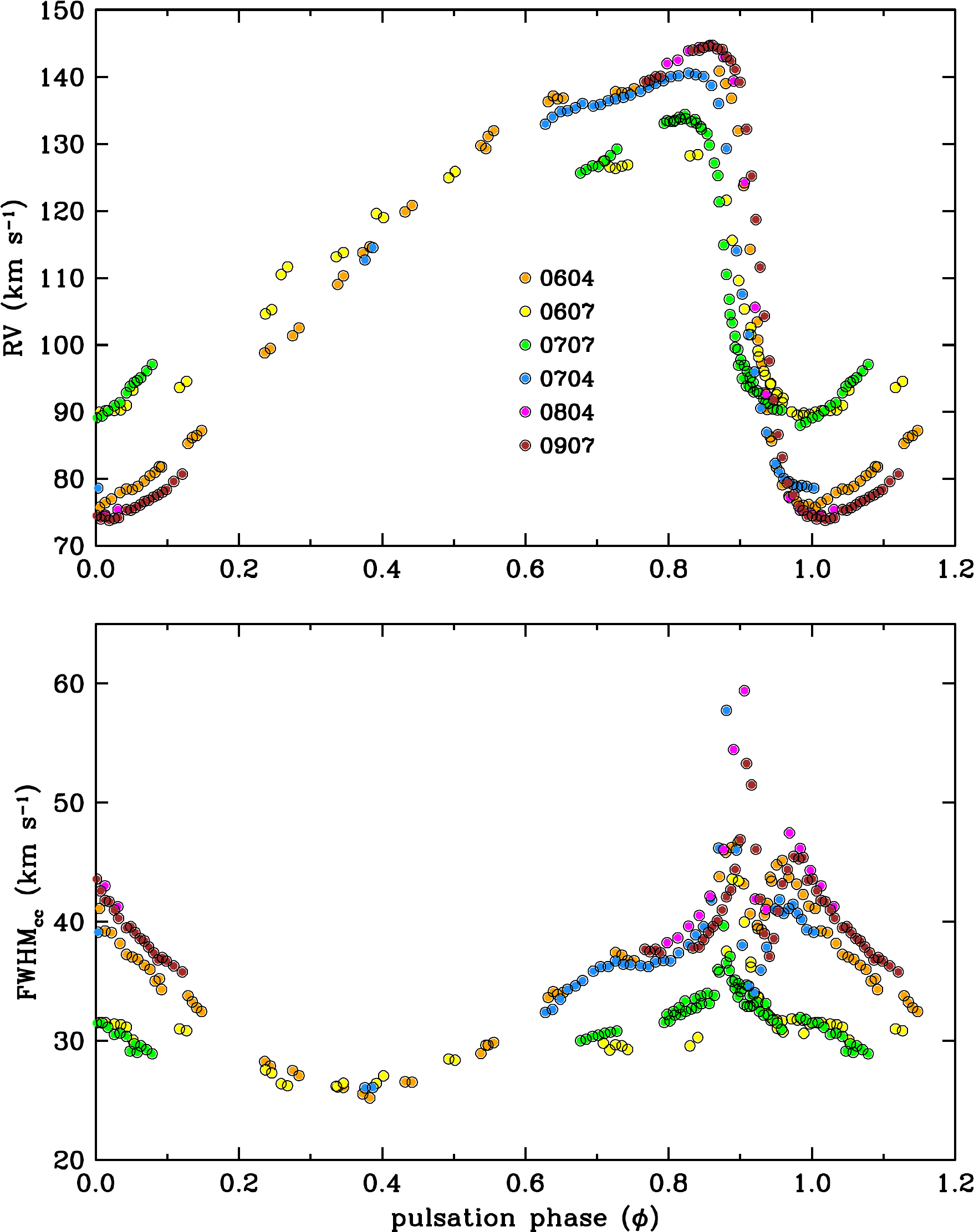}
\caption{\label{fig9}
\footnotesize
   Radial velocity RV (top panel) and $FWHM_{cc}$ of the cross correlation 
   function (bottom panel) versus phase for the RRab Blazhko star UV Oct.  
   Data were obtained by cross correlation with a template spectrum of 
   CS~22874-009 \citep{preston00}.
   The colors indicate different observing seasons, denoted in the figure
   legend as yymm, where yy are the last digits of the year and mm
   are the digits of the month, \eg, 0604 is April of 2006.
}
\end{figure}

\begin{figure}
\epsscale{1.00}
\plotone{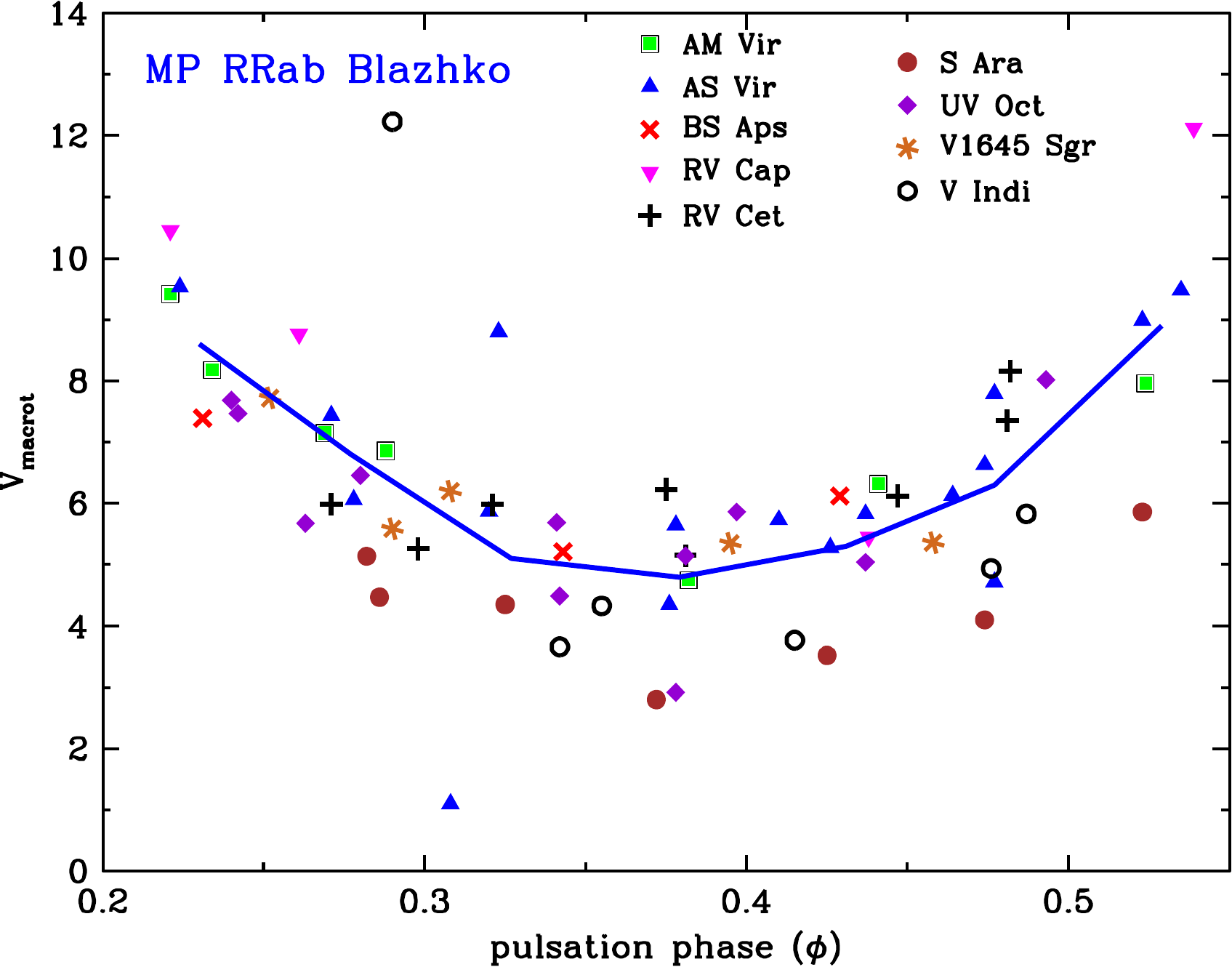}
\caption{\label{fig10}
\footnotesize
   \vmacrot\ versus phase for nine Blazhko RRab stars.  
   The solid black curve denotes mean points calculated for data in 
   successive 0.05 phase intervals.
}
\end{figure}

Complex variations of metallic radial velocities and line widths of Blazhko 
RRab stars near their light maxima (RV minima) defy summary description. 
One example, UV~Oct, is shown in Figure~\ref{fig9}. 
The incomplete sampling of Blazhko phases for UV~Oct is typical for the 
Blazhko stars in our data set.  
Because phases near maximum light contribute nothing to our exploration of 
axial rotation, we restricted our attention to observations made during 
declining light (0.20~$<$~$\phi$~$<$~0.55), where line-widths of stable 
RRab stars achieve their minimum values.  
The ephemerides in Table~\ref{tab3} produce phases of radial 
velocity minima near zero for our Blazhko variables (\citealt{chadid13}, 
Preston, unpublished).  
Values of \vmacrot\ plotted versus these phases are displayed in 
Figure~\ref{fig10}.   
Modest phase corrections, adopted by eye-inspection and listed in column~4 
of Table~\ref{tab3}, were added to phases calculated with these 
ephemerides to reduce horizontal scatter in this figure.
Black line segments connect mean points of observations calculated at 
intervals of ~0.05P.  
The lowest mean point, \vmacrot~=~4.77~$\pm$~0.32~\kmsec, occurring at 
$\phi$~=~0.38, is slightly smaller than the mean value for stable RRab stars, 
$\langle$\vmacrot$\rangle$~=~5.31~$\pm$~0.13~\kmsec. 
Stable RRab and Blazhko RRab stars yield the same upper limit on axial 
rotation of 5~$\pm$1~~\kmsec.

\subsection{An Empirical ``Spectrum of Turbulence'' in RRab Stars}\label{empiricalspec}

\begin{figure}
\epsscale{1.00}
\plotone{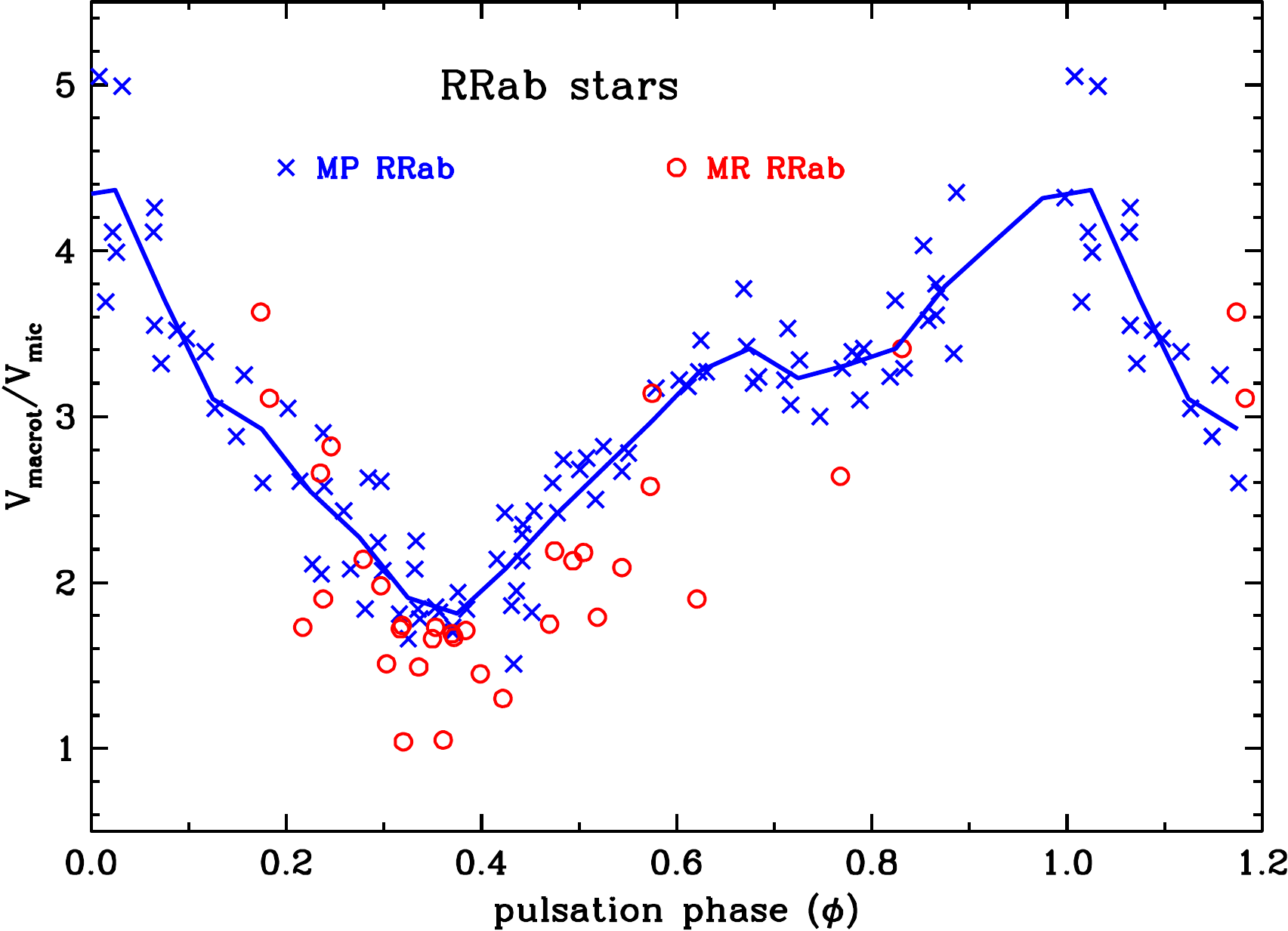}
\caption{\label{fig11}
\footnotesize
   Variation with phase of \vmacrot/\vmic, an empirical spectrum
   of turbulence, for the MP and MR RRab stars.
   The blue curve represents the mean trend for the MP RRab. 
   Its values were computed in phase intervals of $\Delta\phi$~=~0.05;
   see Table~\ref{tab4}.
}
\end{figure}

For stable RRab stars we combine the variations of 
macroturbulence (Figure~\ref{fig7}) and microturbulence 
(Figure~\ref{fig6}) by forming their ratio.
The microturbulent velocities for these stars were derived by \cite{for11b}
and \cite{chadid17} from phase-combined spectra and so there are fewer of
these values than those of \vmacrot.
This is especially true of the \citeauthor{chadid17} results, which were
usually based on only 3 or 4 combined phase points.
We matched phases of \vmacrot\ and \vmic\ data for as many points per
star as possible, interpolating the \vmic\ values where possible and
not attempting any extrapolations into phase intervals not covered by
both data sets.
The results of this exercise are displayed in Figure~\ref{fig11}, where
MP and MR stars are distinguished by different symbols and colors. 
Solid line segments connect mean values of data points calculated in 
phase intervals of 0.05P.  
These mean points are listed in Table~\ref{tab4}.
This variation of the \vmacrot/\vmic\ ratio with phase is in accord with
the minimal requirements of \cite{richardson20}, 
\cite{kolmogorov41}, and the Siedentopf model \citep{woolley53} at all 
phases, namely, that \vmac\ always exceeds \vmic. 
We offer this ratio as a primitive empirical description of a ``spectrum 
of turbulence'', defined by the two length scales associated with macro- 
and micro-turbulence, in RRab atmospheres during their pulsation cycles.

\vspace*{0.2in}
\section{\vmacrot\ FROM EWs and FWHMs OF RRc STARS}\label{RRcRHBvmacrot}

The measurements of EW and \fwhmobs\ for the RRc stars and the conversion 
of these measurements to estimates of \vmacrot\ proceeded in the same manner 
followed for RRab stars in \S\ref{RRabstable}. 
We chose lines for measurement from the same line list. 
Because the RRc stars are systematically hotter than the RRab stars, 
metallic absorption lines are systematically weaker. 
However, the agreement between plateau and regression 
estimates shown in Figure~\ref{fig4} is hardly impaired:
$\langle FWHM({\rm R-P})\rangle$~=~0.46~$\pm$~0.06~\kmsec\ for 
RRc stars, is only 0.11~\kmsec\ larger than the value derived for RRab stars,
0.35~$\pm$~0.05~\kmsec.

\begin{figure}
\epsscale{1.00}
\plotone{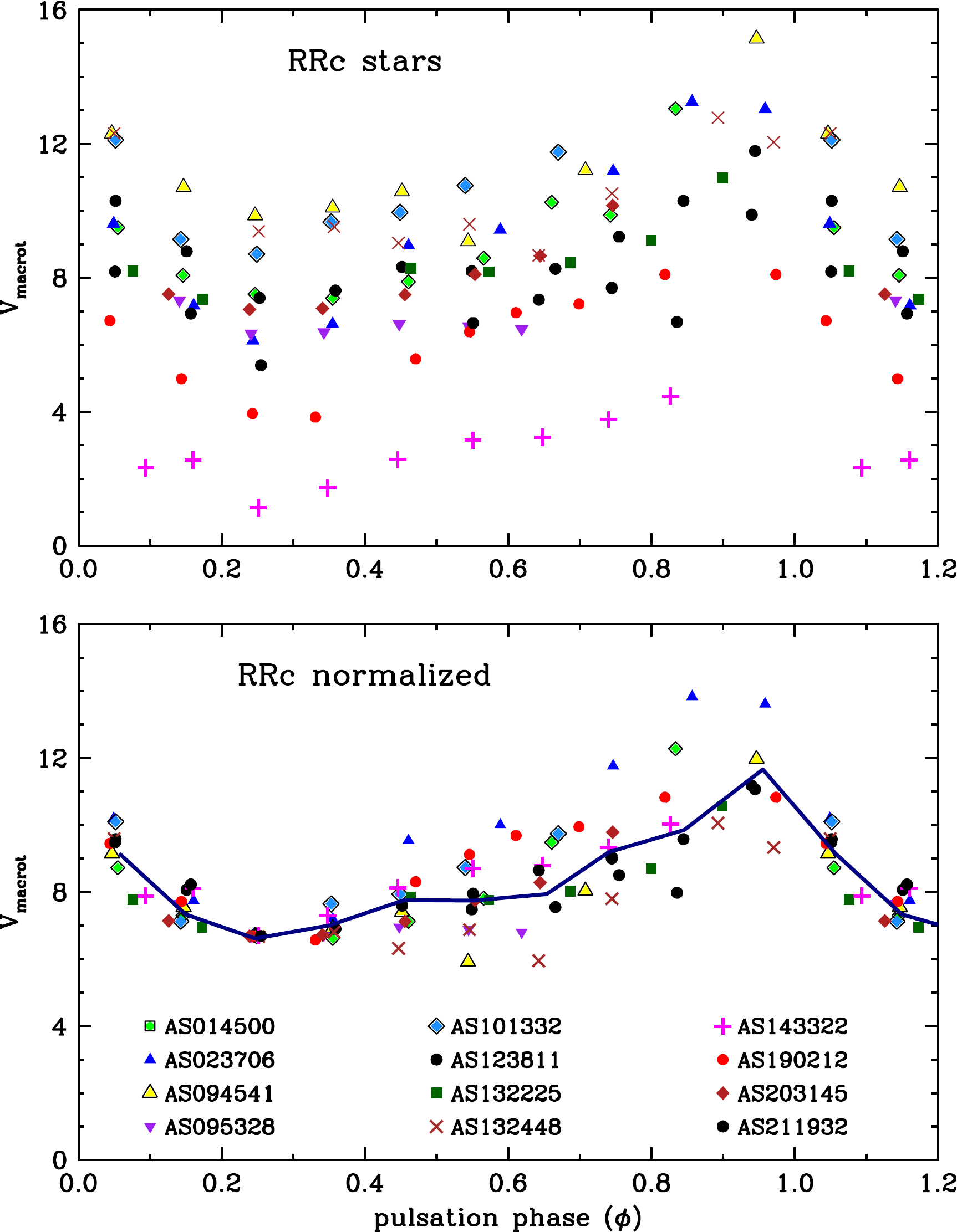}
\caption{\label{fig12}
\footnotesize
   Top panel:  \vmacrot\ versus phase for 12 RRc stars. 
   Bottom panel:  the same data as in the top panel, but with the \vmacrot\
   values in each star arbitrarily shifted to coincide
   with the data point for AS1433 at phase 0.25. 
   The solid curve connects data points phase-averaged in groups of five.
}
\end{figure}

The variations of \vmacrot\ with phase for RRc stars, presented in the 
top panel of Figure~\ref{fig12}, differ markedly from the uniform 
behavior of the RRab stars in Figure~\ref{fig7}.  
Minimum values of \vmacrot\ occur at earlier phases, near phase $\phi$~=~0.3,
and range from 2 to 12~\kmsec.
Maximum values occur near maximum light. 
We create a characteristic variation of \vmacrot\ with phase in the 
bottom panel of Figure~\ref{fig12} by use of additive constants that 
superpose data for all stars near 
phase $\phi$~$\simeq$~0.25 at
\vmacrot~=~6.62~\kmsec, the mean value for the sample.  
This exercise is an instructive, albeit mathematically incorrect, way to 
combine the data. 
The amplitude of the solid black curve that connects mean points of 
individual observations is $\sim$4~\kmsec, much smaller than the amplitude 
for RRab stars ($\sim$10~\kmsec).

\begin{figure}
\epsscale{1.00}
\plotone{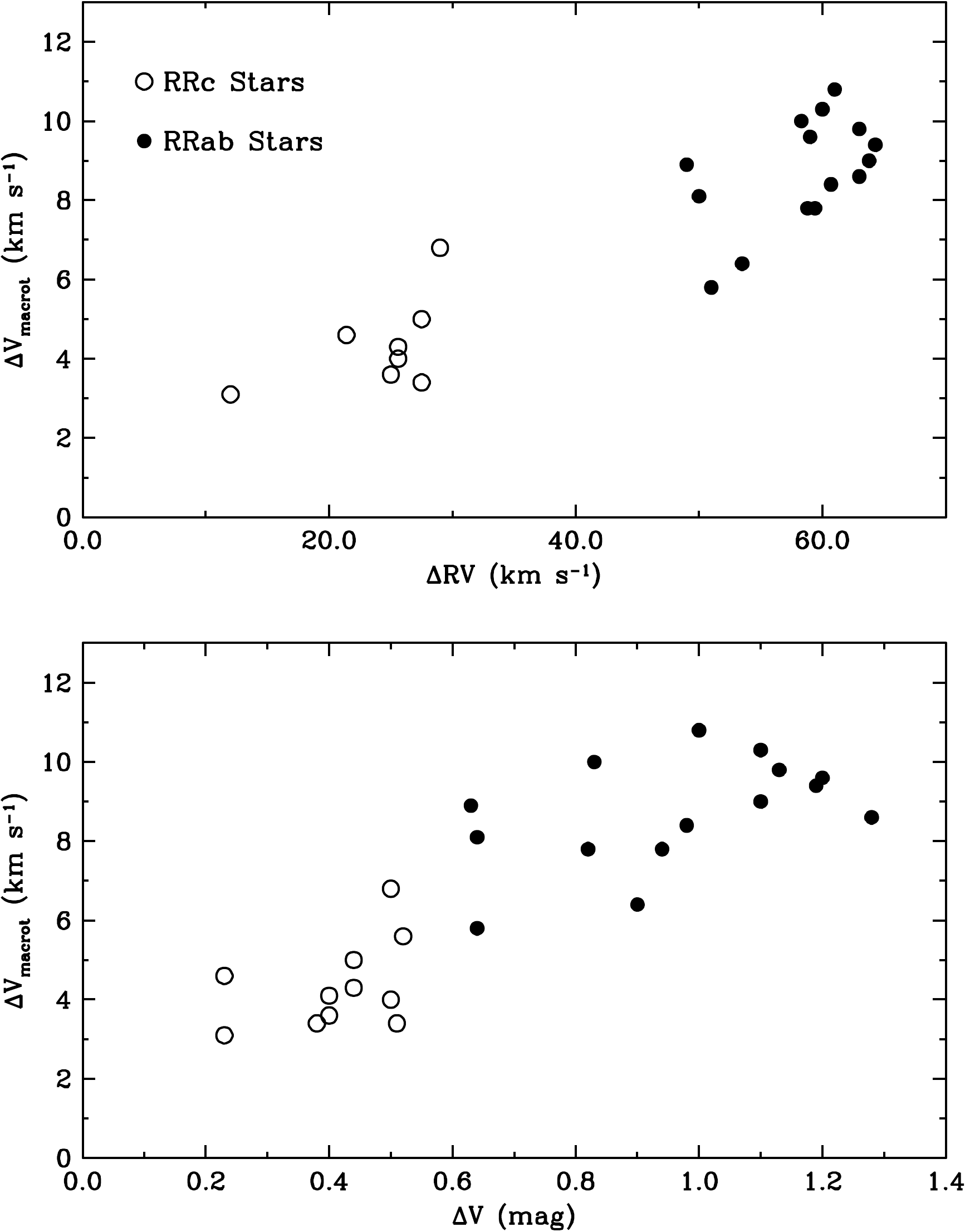}
\caption{\label{fig13}
\footnotesize
   Correlations of $\Delta$\vmacrot\ amplitude (the difference beteween
   maximum and minimum \vmacrot\ values) with (top panel) RV amplitude and 
   (bottom panel) visual light amplitude during pulsation cycles.
}
\end{figure}

We can understand the behavior of the RRc stars in the top panel of 
Figure~\ref{fig12} by noting that macroturbulent velocity amplitude, 
the difference between maximum and minimum values of \vmacrot\ achieved 
in a pulsation cycle, scales with pulsation amplitude for all RR~Lyraes as 
shown by the regressions in Figure~\ref{fig13}.  
Accordingly, we imagine that less macroturbulence in RRc stars with their 
smaller pulsation amplitudes permits detection of smaller axial rotations 
in RRc stars than those encountered among the RRab. 
AS190212-4639 (4~\kmsec) and AS143322-0418 ($\sim$1~\kmsec) are outstanding
examples.
The large upper limit of \vmacrot\ ($\sim$12~\kmsec) for RRc suggests that
BHB rapid rotators adjacent to the instability strip lose their surface angular 
momentum during evolution through the RRc portion of the instability strip,
a possibliity worthy of further exploration with better data.

\vspace*{0.2in}
\section{\vmacrot\ FROM EWs and FWHMs OF METAL-POOR RHB STARS}\label{RHBvmacrot}

We used Magellan MIKE echelle spectra (outlined in \S\ref{obsproc}) 
to estimate \vmacrot\ for twenty-two MP red horizontal branch stars.  
We adopted \fwhminst~=~7.20~\kmsec\ derived from ThAr lines to remove 
instrumental broadening, and we borrowed \vth\ and \vmic\ data from 
\cite{preston06} to calculate values of \vmacrot. 

\begin{figure}
\epsscale{1.00}
\plotone{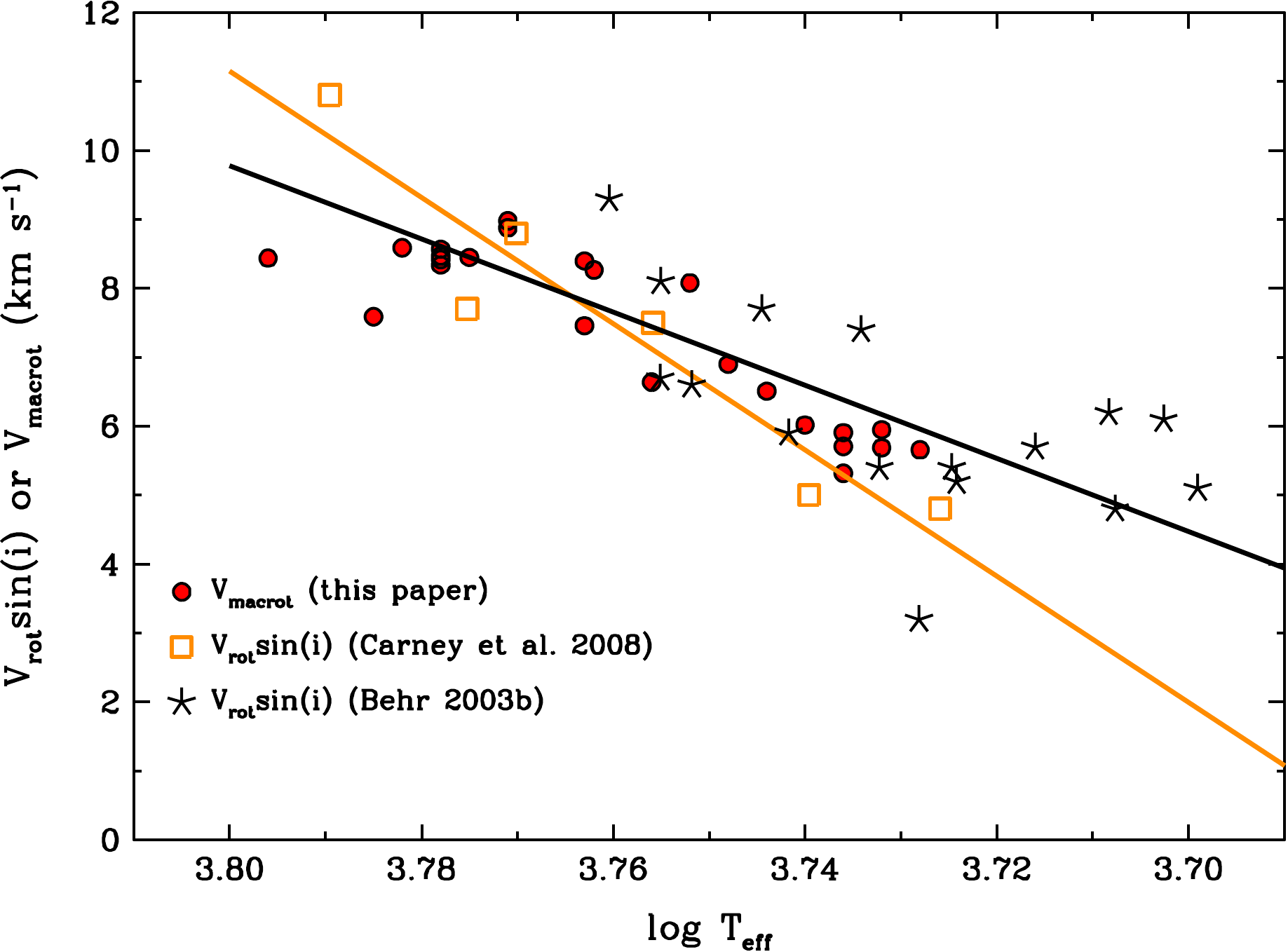}
\caption{\label{fig14}
\footnotesize
   \vmacrot\ values for RHB stars of this study and the $V$sin$i$ values of 
   \cite{behr03b} and \cite{carney08} are plotted versus log \teff.
}
\end{figure}

These \vmacrot\ values are systematically  1 to 2 \kmsec\ larger than 
those of the RRab stars and they decline smoothly with decreasing 
temperature along the RHB portion of the horizontal branch. 
This decline is shared by the values of \vrot\ derived by \cite{behr03b}
and \cite{carney08} as illustrated in Figure~\ref{fig14}.
\cite{behr03b} removed macro-turbulent broadening from his measured line 
widths by the conservative assumption that macro-turbulent velocities equal 
the micro-turbulent velocities derived from his spectrum analyses.  
The mean value and standard deviation of these values for the 
\citeauthor{behr03b} sample presented here are 2.11~$\pm$~0.44~\kmsec. 
\cite{carney08} derived much larger macro-turbulent velocities 
(6 to 12~\kmsec) from Fourier transform analyses of line profiles 
provided by high resolution, high S/N spectra of six RHB stars.  
\citeauthor{carney08} remark that ``rotation and macro-turbulence play 
comparable roles in the line broadening of the observed RHB stars'', 
whereas \citeauthor{behr03b}, by virtue of his assumed small 
macro-turbulent velocity, found that rotation dominates in his sample.  
The data assembled in Figure~3 of \cite{grassitelli15} support the 
conclusions of \citeauthor{carney08}  
Comparison of the only two stars, HD~25532 and HD~184266, common to the 
Behr and Carney et al. samples, are presented in the Table~\ref{tab5}.
We suspect that the larger \vrot\ values of Behr in this table arise, at 
least in part, from his assumed low macro-turbulent velocities.  
Because our \vmacrot\ values lie comfortably between/along the regression 
lines of Carney et al. and Behr in Figure~\ref{fig14} we 
conclude that our \vmacrot\ values are very nearly \vrot\ values, 
\ie, the macro-turbulent velocities of our RHB stars lie below our 
limit of detectability.

We omitted HD~195636 (\vrot~=~20.6~\kmsec, \teff~=~5399~K, 
[Fe/H] = $-$2.74; \citealt{behr03b}) from Figure~\ref{fig14} because 
it lies so far from the locus of all other metal-poor RHB.  
The star is an indubitable metal-poor, giant, rapid rotator (see 
discussion by Preston 1997).

\begin{figure}
\epsscale{0.80}
\plotone{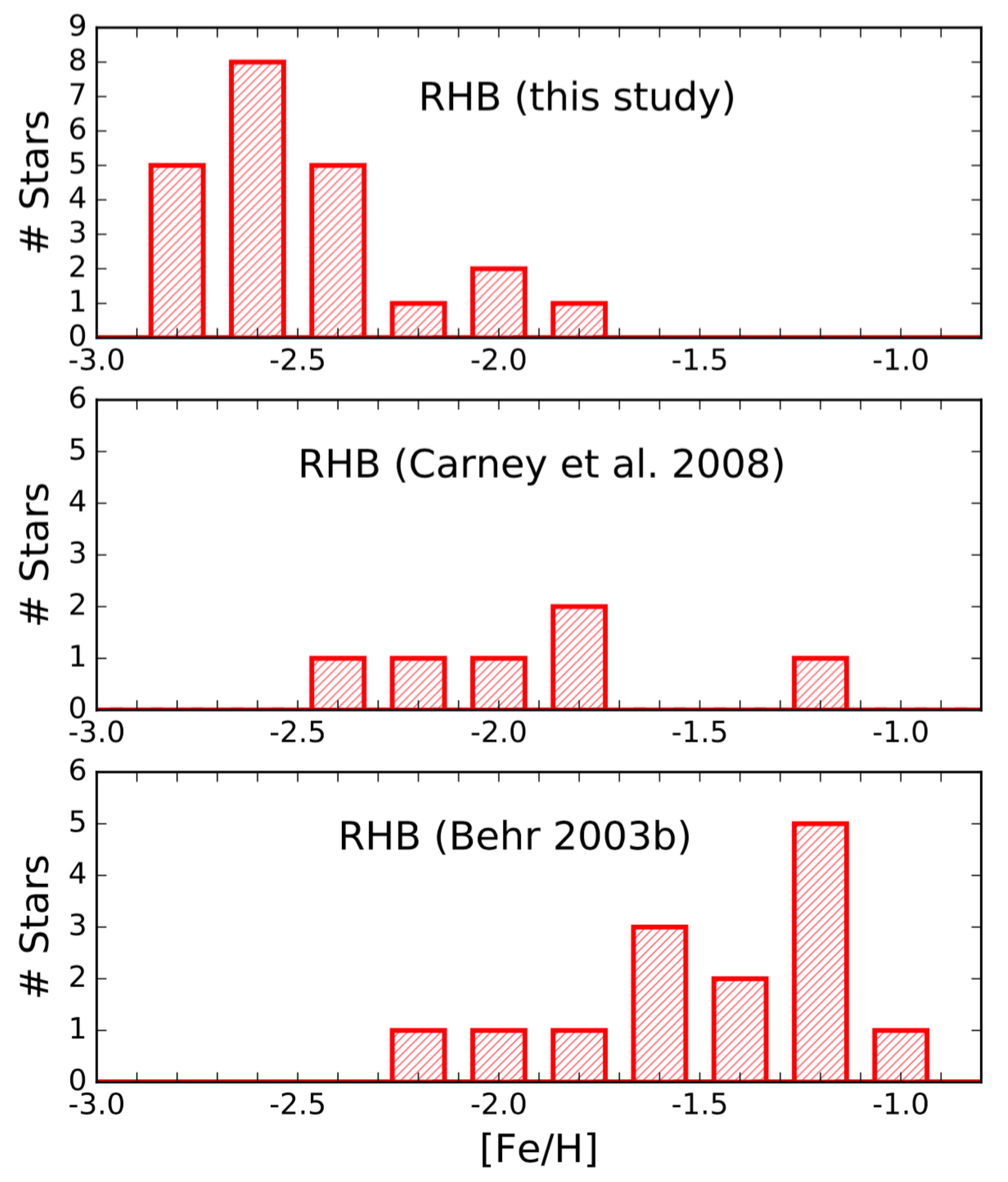}
\caption{\label{fig15}
\footnotesize
   Histograms of the [Fe/H] distributions of three metal-poor RHB samples.
   Strong selection effects are evident.
}
\end{figure}

We seek an explanation for the small values of macro-turbulent 
velocities in our RHB sample in the work of \cite{tanner13} and 
\cite{grassitelli16}, who report that at a given effective temperature the 
optically thin super-adiabatic layers of model atmospheres with higher 
metallicity possess larger convective velocities. 
This is an abundance effect: higher atmospheric opacities produce optically 
thin layers in regions of lower gas density. 
Their calculations induced us to construct the metallicity distributions 
of the three RHB samples, which we display as histograms in 
Figure~\ref{fig15}.  
It is evident at a glance that major selection effects are present.  
The RHB stars of the present study were discovered in the course of
high resolution spectroscopic investigations \citep{preston06,roederer14} 
of the most metal-poor stars, [Fe/H]~$\leq$~$-$2, identified in the survey of 
\cite{beers92}, while the samples of \cite{behr03b} and \cite{carney08} 
were chosen by exploration 
of more abundance-inclusive, photometrically-defined samples. 
The content of the histograms is summarized in 
Table~\ref{tab6}.

The mean [Fe/H] values of this study and that of \cite{behr03b} differ 
by one full order-of-magnitude. 
The sample fractions with [Fe/H]~$>$~$-$1.7 in these three studies 
range from 0.0 to 0.8.  
This comparison of our very metal-poor sample with that of \cite{carney08}
supports the conclusion of \cite{tanner13}: macro-turbulent velocities 
in the optically thin layers that produce metal absorption lines 
increase with increasing metallicity. 
Perhaps this presentation will stimulate renewed interest in 
abundance-dependent macro-turbulence on the horizontal branch.

\vspace*{0.2in}
\section{DISCUSSION}\label{summary}

\begin{figure}
\epsscale{1.00}
\plotone{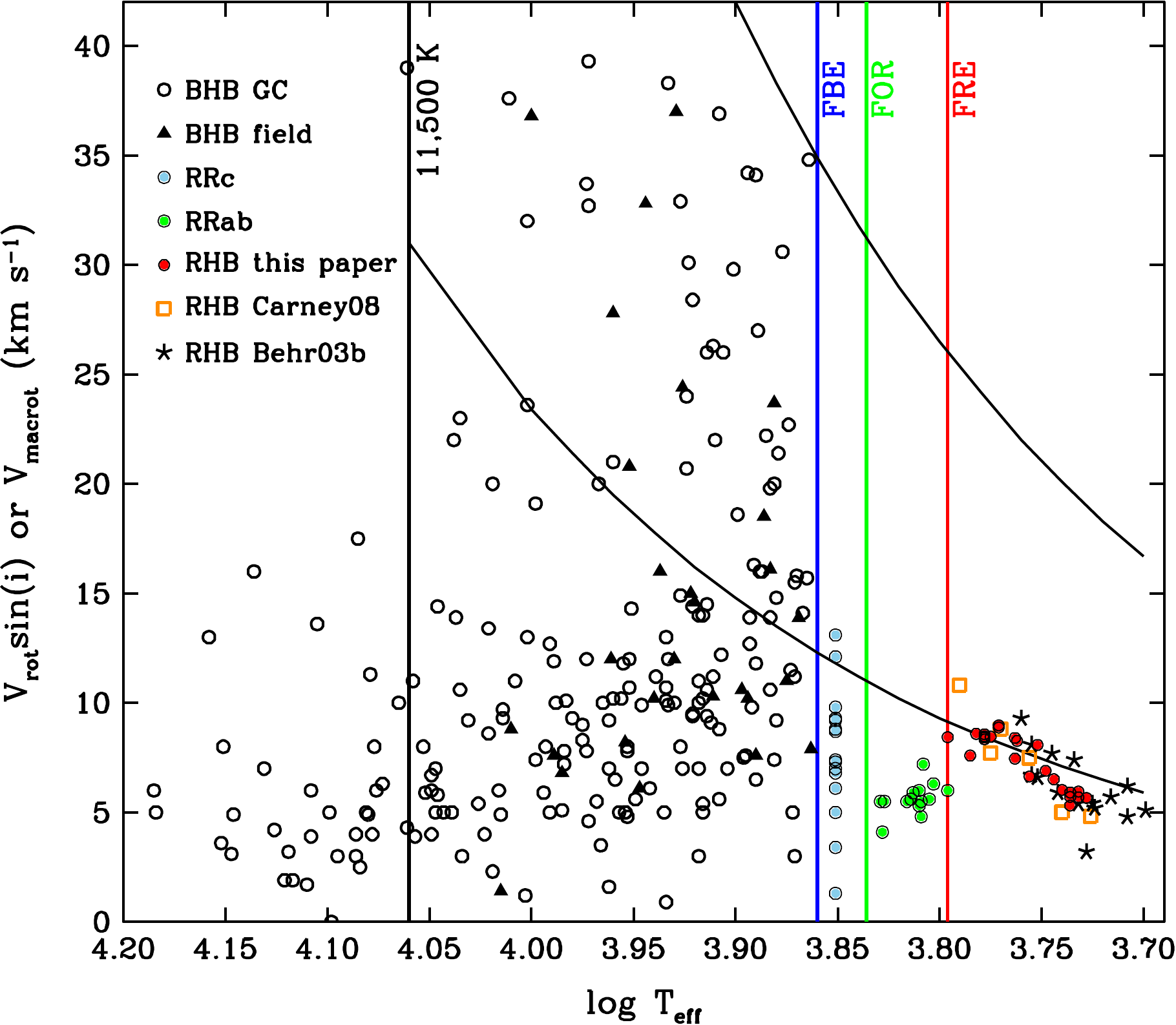}
\caption{\label{fig16}
\footnotesize 
   Variation of \vrot\ (BHB stars) and \vmacrot\ (RR Lyrae and RHB stars) with 
   log(\teff) along the metal-poor horizontal branch.
   The black vertical line denotes the Grundahl limit at 11,500~K.  
   For the instability strip the vertical lines are: the fundamental blue 
   edge (FBE, blue); the first overtone red edge (FOR, green), and the 
   fundamental red edge (FRE, red).
   Black parabolic trajectories define the regions of the instability strip
   traversed by BHB rapid rotators that evolve with constant Luminosity
   and constant angular momentum ($V_{rot}$~$\propto$~\teff$^2$).
}
\end{figure}

In this paper we have combined high resolution spectra 
of RR~Lyrae and RHB stars collected at Las Campanas Observatory over three 
decades with literature studies of RHB and BHB stars to derive rotation 
rates or meaningful upper limits for stars along the horizontal branch.
In Figure~\ref{fig16} we present a summary plot, showing \vmacrot\
or \vrot\ as a function of effective temperature along the HB.
The astonishing variations of measured rotational velocity with \teff\
clearly indicate that surface angular momentum cannot be a proper measure of 
total angular momentum of stars along the HB.

To begin discussion of Figure~\ref{fig16}, we recall that
the large values of \vrot\ for many metal-poor BHB stars first 
encountered by \cite{peterson95} at \teff~$<$~11500~K
were totally unexpected in view of the small projected rotational 
velocities (\vrot~$<$~5~\kmsec) of their evolutionary antecedents 
in globular clusters \citep{lucatello03}.  
Then, the abrupt decline in rotation at the blue edge of the instability 
strip \citep{peterson96} was an additional surprise, appropriately 
labeled a ``conundrum'' by \cite{preston11} and \cite{preston13}.  
This investigation reduces the upper limit of RRab rotation 
from 10 to 5~\kmsec, and additionally suggests that rapid BHB rotation 
disappears during RRc evolution. 
Should the arguments of \cite{tanner13} apply to RR~Lyrae, the upper limit on 
RR~Lyrae rotation would be further reduced.  

\begin{figure}
\epsscale{0.80}
\plotone{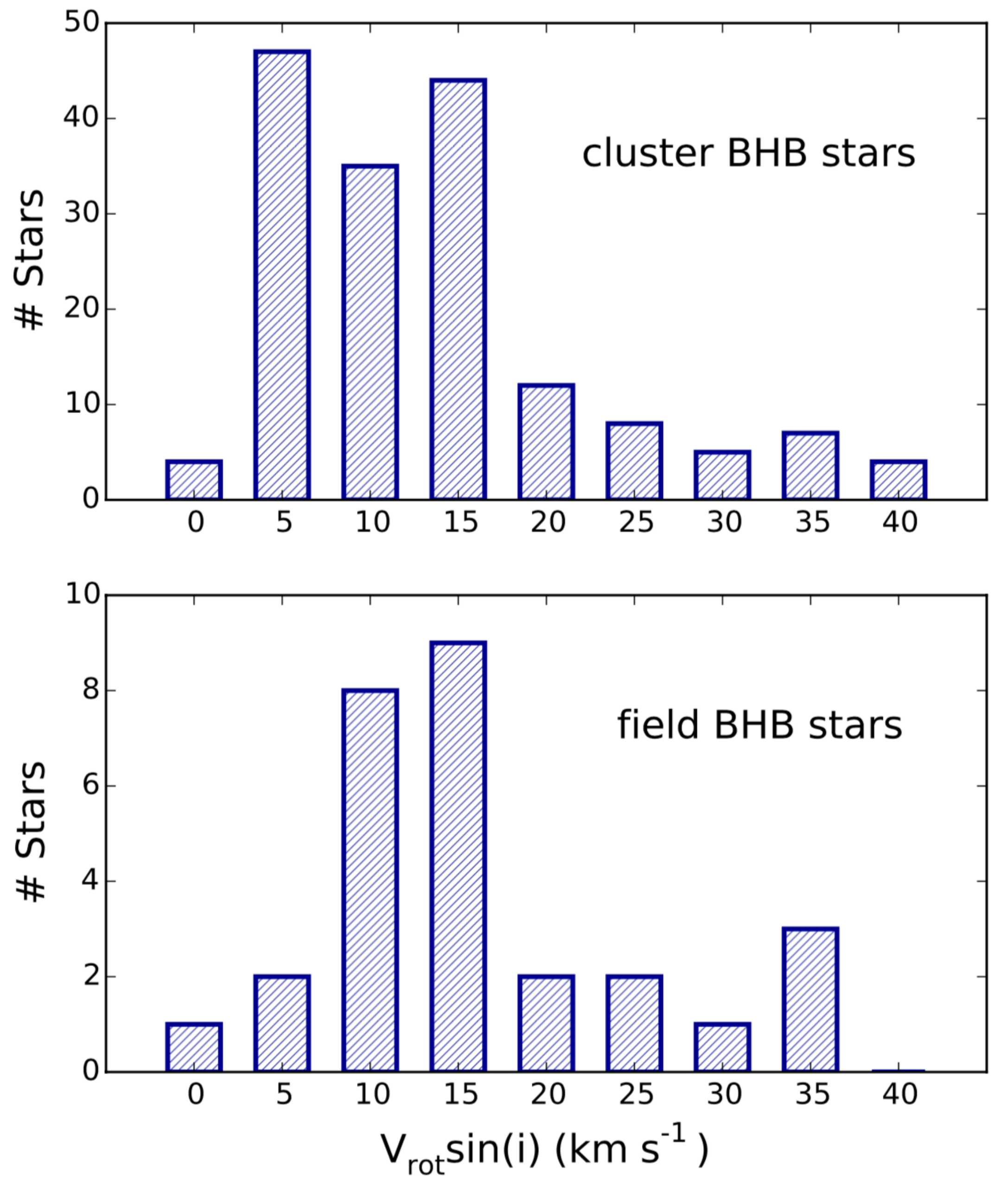}
\caption{\label{fig17}
\footnotesize
   The \vrot\ distributions of globular cluster (top panel) and field 
   (bottom panel) BHB stars in the compilation of \cite{cortes09}.
}
\end{figure}

Extant \vrot\ distributions differ from cluster to 
cluster in the Galactic halo.  
However, whether field and globular cluster HB star samples are drawn 
from the same parent population is an open question.  
The \vrot\ distributions of the \cite{cortes09} compilations of field 
and cluster HB stars are similar, as noted previously by \citeauthor{cortes09}
and shown in Figure~\ref{fig17}.  
This similarity encouraged us to include the cluster HB stars in 
Figure~\ref{fig16}.  
Only a handful of the $\sim$150 known Galactic globular clusters have been 
investigated for rotation, and the field sample is limited to HB stars near the
solar circle, so it is premature to draw firm conclusions about this issue.

\begin{figure}
\epsscale{1.00}
\plotone{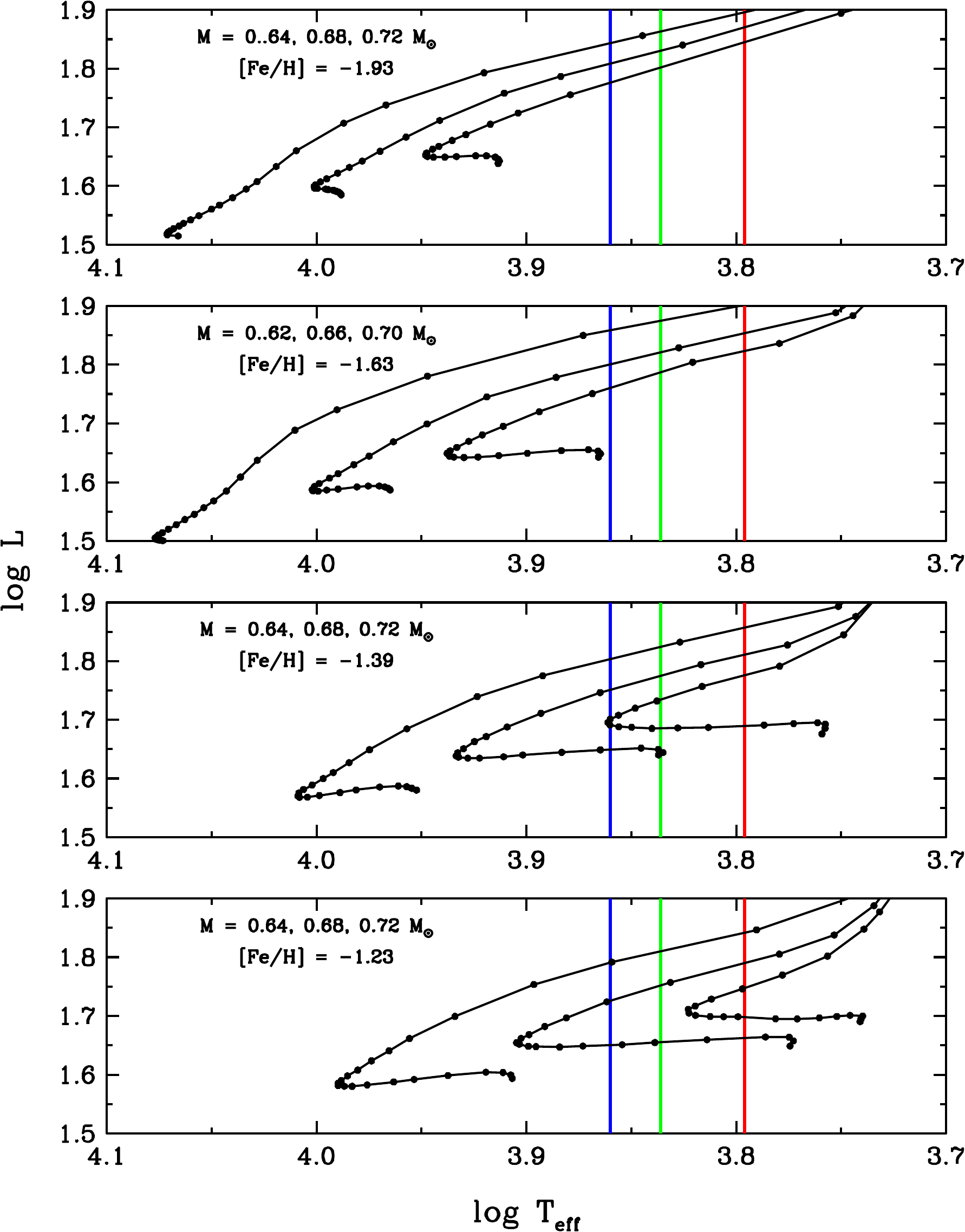}
\caption{\label{fig18}
\footnotesize
   Evolutionary tracks for HB stars with masses and abundances that 
   bracket the models adopted by \cite{vandenberg16} for the globular 
   clusters M5, M15, and M92. 
   Blue, green, and red vertical lines mark the fundamental blue edge, first 
   overtone red edge, and fundamental red edge of the instability strip.
}
\end{figure}

The use of ``during'' in the description of RRc evolution deserves 
comment because of the complicated nature of HB evolution in and near 
the instability strip.  
This is best appreciated by inspection of the evolutionary tracks of 
\cite{lee90} in Figure~\ref{fig18} plotted for Y~=~0.23 with masses and 
abundances that bracket the models adopted for the metal-poor globular 
clusters M3, M15, and M92 by \cite{vandenberg16}.  
For nearly all of the models plotted in Figure~\ref{fig18}, approximately 
horizontal evolution at near-constant Luminosity
toward higher effective temperatures occupies $\sim$2/3 of their 
horizontal branch lifetimes (see Figure~1 of \cite{desantis99}).  

\begin{figure}
\epsscale{0.80}
\plotone{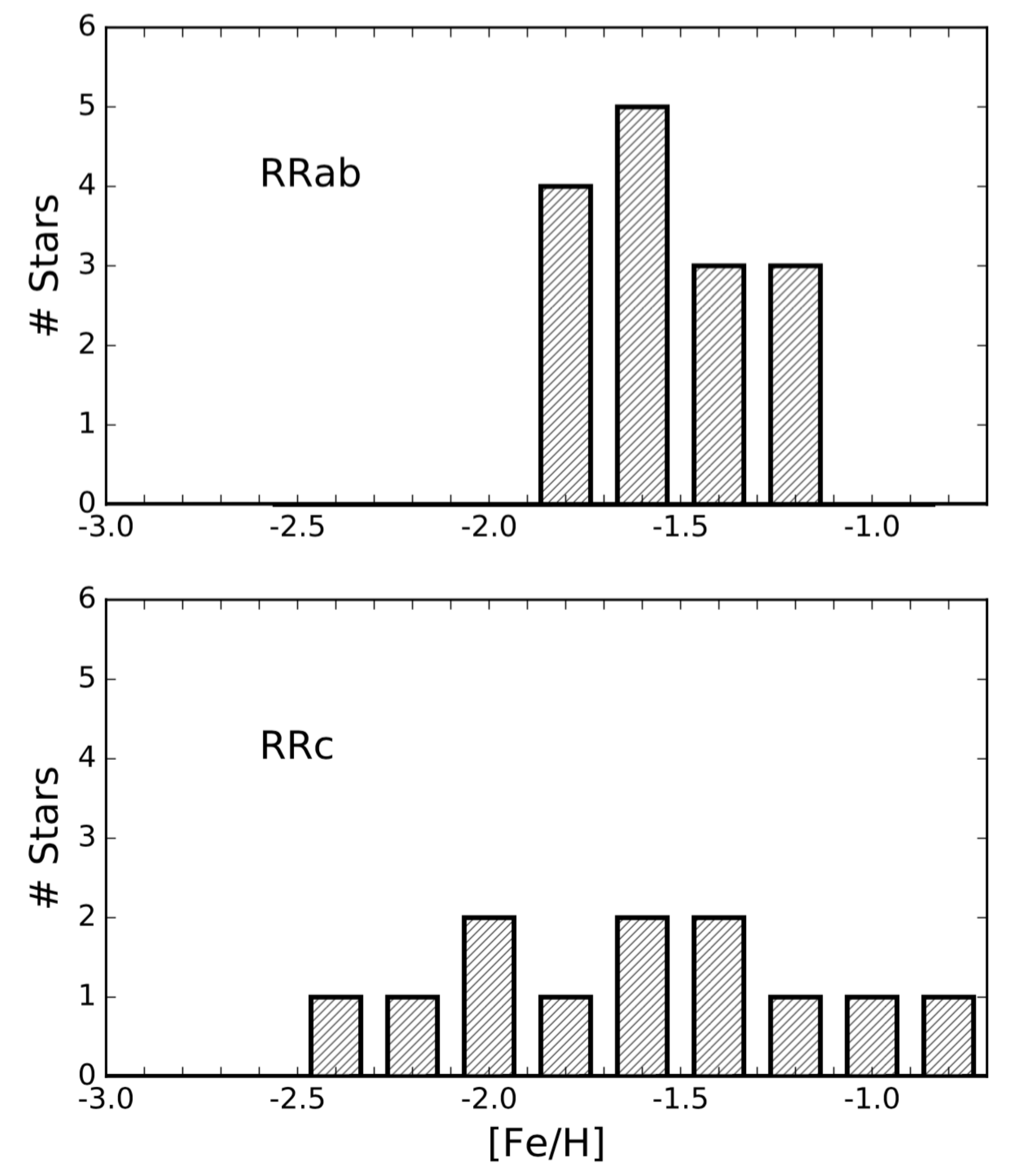}
\caption{\label{fig19}
\footnotesize
   Abundance distributions of the RR Lyrae stars studied in this paper.
}
\end{figure}

Second parameter issues play no role in our study. 
Evolution from the Zero Age Horizontal Branch (ZAHB) to higher temperature,
the ``blue loops'' in 
Figure~\ref{fig18} is the result of a hydrogen-burning phenomenon 
common to all HB evolutionary sequences
(\citealt{sweigart87}, \citealt{lee90}, \citealt{castellani91}, 
\citealt{dorman92}, \citealt{dotter07}) for reasons discussed in some detail 
by \citeauthor{sweigart87},  by \citeauthor{dorman92}, and by 
\citeauthor{dotter07}.
Early on \cite{rood73} realized that these loops could not provide the range 
in HB color distributions required of a second parameter.  
However, the blue loops would act to blur the pattern of \vmacrot\ values 
of RHB and RR~Lyrae stars seen in Figure~\ref{fig16}, were
ZAHB stars to retain their initial surface angular momenta during 
migration from the RHB and/or RR~Lyrae domains toward the BHB.
For [Fe/H]~$<$~$-$1.6 evolution proceeds steadily from the BHB 
through the instability strip to the RHB and on to the AGB.  
However, for stars with [Fe/H]~$>$~$-$1.6, evolution can be 
quite complicated.  
As a consequence of stochastic mass loss, an RRc star-to-be may 
arrive on the ZAHB variously as an RHB, 
RRab, RRc or BHB star, all of these evolving to higher temperatures 
for most of their HB lifetimes.  
From the metallicity distributions of our RR~Lyrae samples shown in 
Figure~\ref{fig19} we see that roughly half of them, those with 
[Fe/H]~$<$~$-$1.6, evolve monotonically from the BHB through the 
instability strip to the AGB.  
At higher metallicity a star can arrive on the ZAHB as an RHB star, 
evolve through the RRab and RRc states to the BHB and then go back 
across the instability strip to the AGB.  
However, Figure~\ref{fig19} indicates that the stars retain no memory 
of their initial ZAHB locations.  

We observe a decline in mean \vrot\ during the transition from 
BHB to RRc; then a gentle rise in the transition from RRab to RHB, 
followed by a decline during subsequent RHB evolution toward the AGB.  
Absent a mechanism for altering the total angular momentum of HB stars, 
we attribute these regularities in behavior of \vrot\ among the RR~Lyrae 
and RHB stars as indications that angular momentum is continuously 
being redistributed within these stars on time scales short, 
$\sim$10$^7$~y, compared to HB lifetimes.  
\cite{sills00} discuss schemes for such redistribution.

As regards removal of surface angular momentum that might accompany 
mass loss during RR Lyrae evolution, the only credible investigation of which 
we are aware is the 3D numerical simulation of \cite{stellingwerf13}, 
which produced a hot corona with outflow.  
Stellingwerf did not estimate the mass loss rate, and we are loathe to 
speculate in this matter.  
Whether or not these notions are correct, we believe that they are 
legitimate food for thought worthy of presentation to theoreticians.

Finally, we note the inclusion in Figure~\ref{fig16} of HD~195636 
with log~\teff =  3.732, \vrot~=~20.6~\kmsec, [Fe/H]~=~$-$2.7 \citep{behr03b} 
lying far above all other RHB stars.
We deliberately excluded this star from the discussion of \S\ref{RHBvmacrot} 
simply because it is such an egregious outrider.  
A long history of investigations, summarized by \cite{preston97}, confirm 
that it is a very metal-poor RHB or AGB star with lines abnormally broad 
for its presumed evolutionary state.  
Further speculation about the history of HD~195636 is beyond the scope 
of this paper.

\acknowledgments

We thank all the Las Campanas Observatory support personnel for their help 
during the course of our endeavor, with particular regards to several 
du Pont telescope operators for their efforts in assisting with the 
observations required to produce this paper.  
We also thank Michel Breger, Marcio Catelan, Bill Cochran, and Phillip 
MacQueen for helpful comments that improved our presentation.
This work has been supported in part by NSF grant AST1616040 to C.S.  and by 
the University of Texas Rex G. Baker, Jr. Centennial Research Endowment.

\appendix
\section{Estimates of \fwhminst\ of the duPont echelle spectrograph}

We explore two ways to estimate the instrumental width, \fwhminst, of 
the duPont echelle spectrograph.

\begin{figure}
\epsscale{1.00}
\plotone{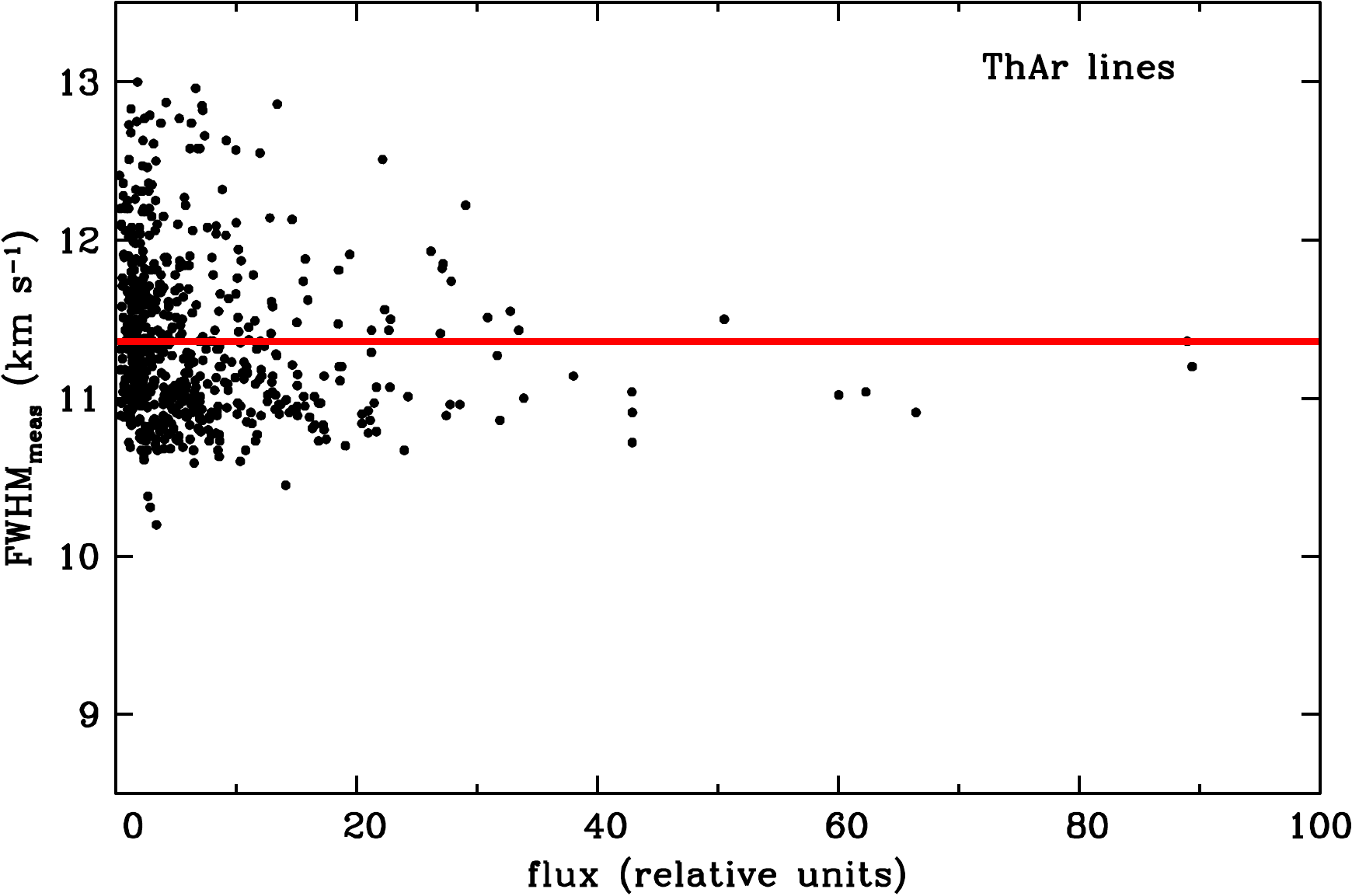}
\caption{\label{fig20}
\footnotesize
   $FWHM$ of ThAr emission lines plotted versus the relative integrated
   line fluxes.
}
\end{figure}

(1)  Using procedures described in \S\ref{instcorr}, we made 591 
measurements of $EW$ (arbitrary flux units) and $FWHM$ (\kmsec) in six 
echelle orders of fourteen ThAr spectra chosen at random from observations 
made during observing runs in the years 2006, 2007, 2009, 2011, and 2012.  
The results are plotted in Figure~\ref{fig20} as a plot of $FWHM$ versus 
$EW$, where the horizontal line represents the average value, 
11.36~$\pm$~0.52~\kmsec, of all the measurements, 
The average and standard deviation for the 151 strongest lines 
are reduced only slightly to 11.25~$\pm$~0.45~\kmsec.
Inspection of Figure~\ref{fig20} reveals an obvious skew toward larger 
$FWHM$ values.
This asymmetry arises from degradation of the echelle instrumental profile 
with increasing angular distance from the optical axis of the spectrograph.  
This degradation is due to optical aberrations and to inevitable slight 
tilt of the CCD chip with respect to the focal surface of the spectrograph 
camera.

\begin{figure}
\epsscale{1.00}
\plotone{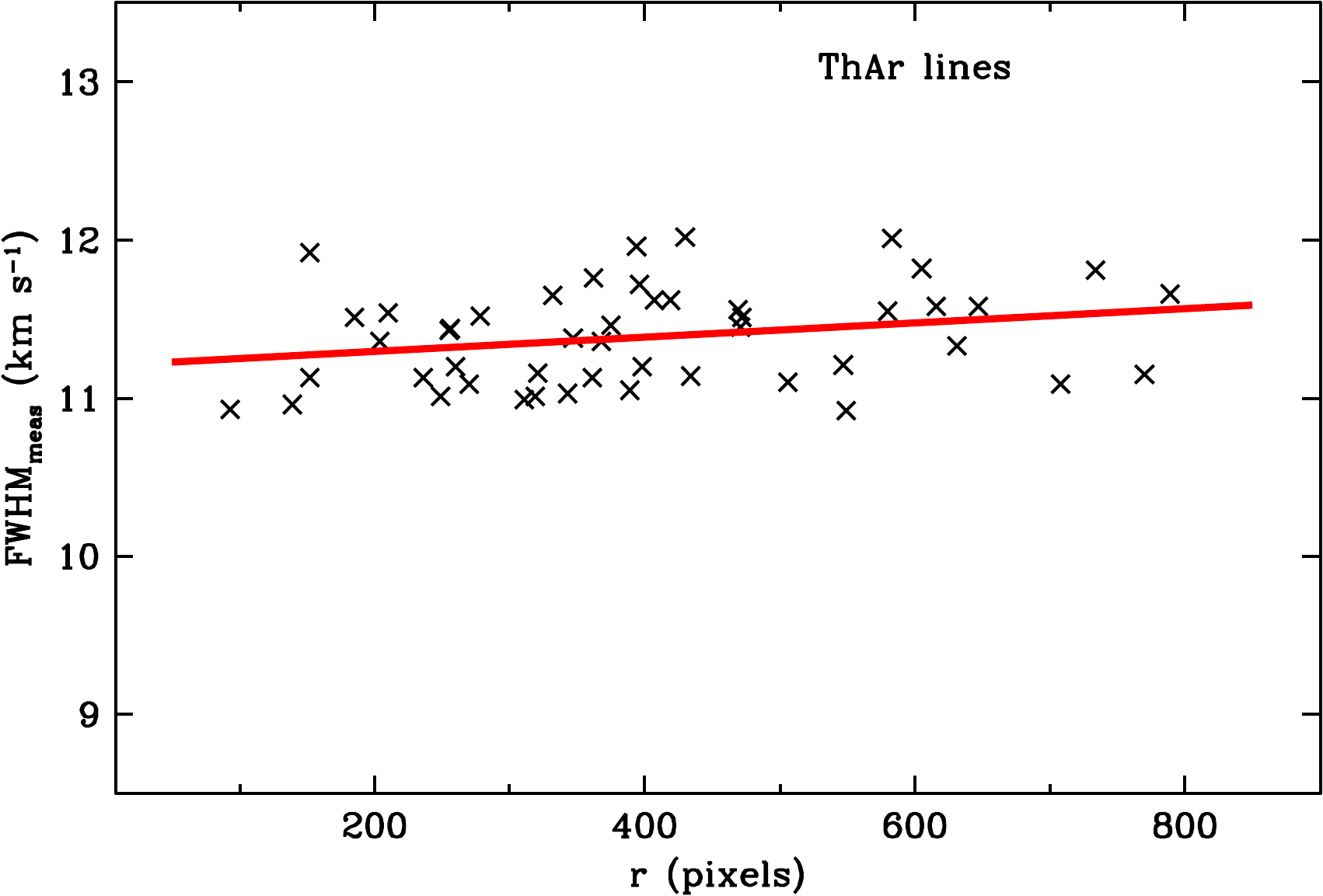}
\caption{\label{fig21}
\footnotesize
   $FWHM$ of ThAr lines plotted versus radial pixel distance from the 
   center of the CCD chip. 
   The red line is a linear regression derived from the plotted data points.
}
\end{figure}

We tested the hypothesis that skew is introduced by position-dependent 
aberrations by using the (x,y) pixel coordinates of ThAr the lines in the 
CCD image frames to calculate the radial distance r(x, y) of each line from 
the central coordinates of the CCD chip (x=1024, y=1024).  
The results of this exercise are shown in Figure~\ref{fig21}.
We found that $\langle FWHM\rangle$ increases with 
r(x, y) by $\sim$0.4~\kmsec\ over the usable range of r(x, y), 
in qualitative agreement with expectations.

\begin{figure}
\epsscale{1.00}
\plotone{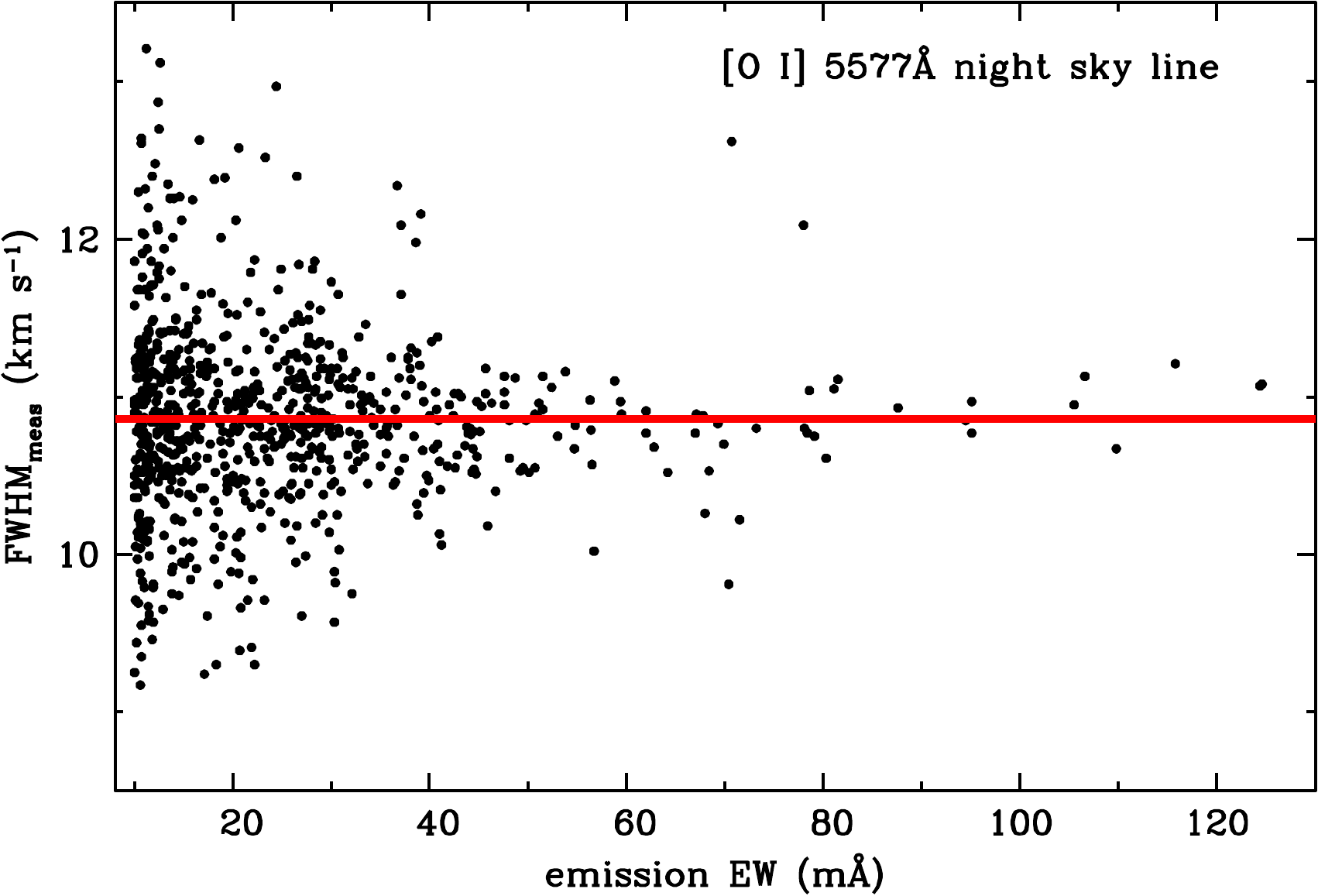}
\caption{\label{fig22}
\footnotesize
   \fwhmobs\ of the night sky [\species{O}{i}] 5577 Å emission
   line plotted versus its $EW$ in units of normalized stellar fluxes.
}
\end{figure}

(2) Following the suggestion of an anonymous referee, we measured 
$FWHM$ of the airglow line [\species{O}{i}] 5577~\AA\ in 792 spectra of 
RR~Lyrae stars observed in the years 2006$-$2012.  
This single telluric line is not ideal for our purposes, because it lies at 
a single location on the CCD chip some 1000~\AA\ (20 echelle orders)
longward of the bulk of our ThAr and stellar line measurements.  
Our measurements are displayed in Figure~\ref{fig22} as a plot of $FWHM$ 
versus $EW$ of the airglow emission in units of normalized stellar continua.  
The [\species{O}{i}] emission feature is characteristically weak in our 
spectra because of our short ($<$600~s) exposure times.
The red horizontal line in this figure represents the average of all 
measurements, 10.86~$\pm$~0.62~\kmsec.  
The scatter of individual $FWHM$ about this mean value is free of skew 
because it is located at the same near-central location on the CCD chip 
in all spectra.
The mean $FWHM$ of the [\species{O}{i}] distribution is 0.50~\kmsec\ smaller 
than that of the ThAr lines.
No reasonable assumptions about thermal broadening of [\species{O}{i}]
5577~\AA\ in the Earth’s atmosphere or Th lines in hollow cathode lamps 
can produce this difference, the cause of which remains unknown to us.

\begin{figure}
\epsscale{1.00}
\plotone{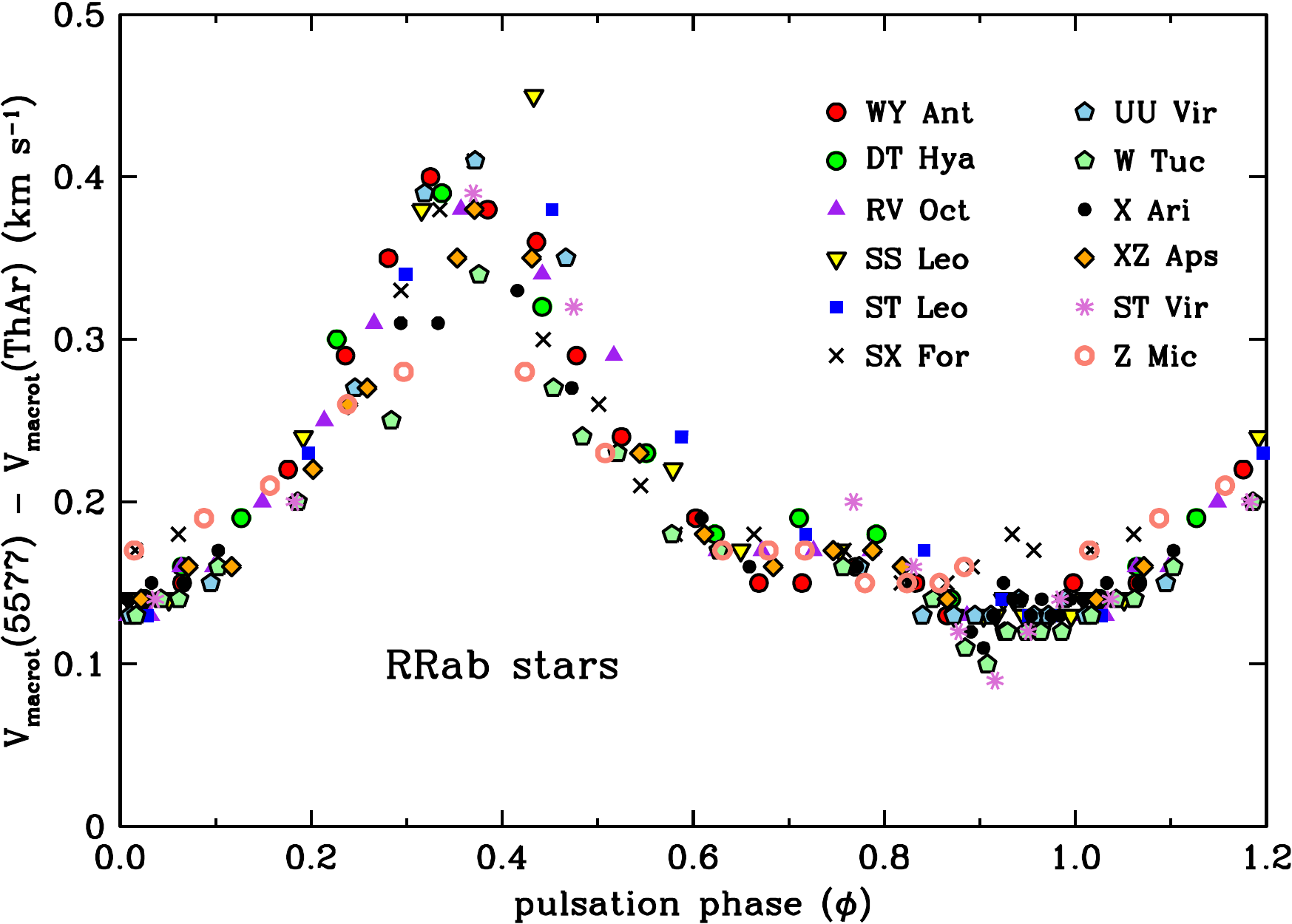}
\caption{\label{fig23}
\footnotesize
   The increases in \vmacrot, if \fwhminst\ is reduced from 11.36~\kmsec\ to 
   10.86~\kmsec, are plotted as a function of pulsation phase for 12 RRab stars.
}
\end{figure}

We consider how our conclusions about \vmacrot\ depend on our 
choice of \fwhminst\ by calculating the difference in 
\vmacrot(5577)~$-$~\vmacrot(ThAr) as a function of phase. 
In Figure~\ref{fig23} we display these differences 
for the twelve RRab stars identified in the figure legend.
In this exercise we used 10.86~\kmsec, the average for all measurements of 
[\species{O}{i}] 5577~\AA, and 11.36~\kmsec, the average for all 
measurements of ThAr lines.  
The smaller \fwhminst\ derived from [\species{O}{i}]5577~\AA\ produces 
larger \vmacrot\ values.  
These differences in \vmacrot, which vary with pulsation phase are generally 
smaller than 0.4~kmsec\ near phase $\phi$~=~0.38, the phase we use to set 
the upper limit on \vrot.
This difference, 0.4~\kmsec, provides an independent estimate of uncertainty 
in our upper bound of 5~\kmsec\ for the \vrot\ of metal-poor RRab stars.  
We prefer the larger $FWHM$ derived from the ThAr lines, because these 
lines are distributed more nearly like those of the metallic absorption 
lines which occur at random locations in all spectral orders.

(3) Finally, consider the narrow-lined HD 140283, our radial 
velocity standard.
It has a measured \fwhmobs~= 12.15~\kmsec\ that narrowly exceeds our adopted
\fwhminst~=~11.36~\kmsec.  
We calculate \vmacrot~=~1.6~\kmsec\ for HD 140283 by
use of these FWHM values, \teff~=~5750~K, and \vmic~=~1.50~\kmsec.
\cite{collet09} used their 3D hydrodynamic code to derive 
\vrot~=~2.58 \kmsec\ for HD140283.  
The difference between their value and ours, 1.0~\kmsec, is consistent 
with our estimate of the uncertainty in our calculations, and it is 
unclear which of these two values is to be preferred.

\bibliographystyle{apj}



\clearpage
\startlongtable
\begin{center}
\begin{deluxetable}{cccccccccc}
\tabletypesize{\scriptsize}
\tablewidth{0pt}
\tablecaption{The Stellar Sample\label{tab1}\tablecolumns{10}}
\tablehead{
\colhead{Star Name}                      &
\colhead{[Fe/H]}                         &
\colhead{$P$}                            &
\colhead{$HJD_0$}                        &
\colhead{$V$}                            &
\colhead{log(\teff)\tablenotemark{a}}    &
\colhead{$\langle FWHM_{unsat}\rangle$}  &
\colhead{$\sigma (FWHM_{unsat})$}        &
\colhead{$\langle$\vmacrot$\rangle$}     &
\colhead{$\sigma$(\vmacrot)}             \\
\colhead{}                               &
\colhead{}                               &
\colhead{days}                           &
\colhead{days}                           &
\colhead{max light}                      &
\colhead{(K)}                            &
\colhead{\kmsec}                         &
\colhead{\kmsec}                         &
\colhead{\kmsec}                         &
\colhead{\kmsec}                         
}
\startdata
\multicolumn{10}{c}{MP RRab Calibration} \\
WY Ant         &   $-$1.66 &  0.574344 & 3835.592 & 10.4 &   3.810 & 10.2 &       0.3 &  5.3 &      0.2 \\
XZ Aps         &   $-$1.57 &  0.587266 & 3836.276 & 11.9 &   3.809 & 10.5 &       0.2 &  5.5 &      0.2 \\
SW Aqr         &   $-$1.24 &  0.459303 & 1876.138 & 10.6 &   3.829 & 10.5 &   \nodata &  5.5 &  \nodata \\
X Ari          &   $-$2.50 &  0.651170 & 1890.064 &  9.2 &   3.803 & 11.7 &       0.2 &  6.3 &      0.1 \\
RR Cet         &   $-$1.52 &  0.553029 & 1900.312 &  9.3 &   3.813 & 11.2 &   \nodata &  5.9 &  \nodata \\
SX For         &   $-$1.62 &  0.605342 & 1870.406 & 10.9 &   3.805 & 10.7 &       1.2 &  5.6 &      0.8 \\
VX Her         &   $-$1.60 &  0.455359 & 2699.996 & 10.1 &   3.828 &  8.6 &       0.6 &  4.1 &      0.5 \\
DT Hya         &   $-$1.22 &  0.567978 & 3835.608 & 12.5 &   3.814 & 10.7 &       0.9 &  5.6 &      0.6 \\
SS Leo         &   $-$1.83 &  0.626335 & 1873.056 & 10.5 &   3.809 &  9.5 &       0.6 &  4.8 &      0.4 \\
ST Leo         &   $-$1.29 &  0.477984 & 5322.597 & 11.0 &   3.827 & 10.6 &       0.4 &  5.5 &      0.3 \\
Z Mic          &   $-$1.28 &  0.586928 & 3930.509 & 11.3 &   3.808 & 13.1 &       0.1 &  7.2 &      0.1 \\
RV Oct         &   $-$1.34 &  0.571170 & 3835.895 & 10.5 &   3.816 & 10.6 &       0.4 &  5.5 &      0.3 \\
VY Ser         &   $-$1.82 &  0.714105 & 5325.340 &  9.8 &   3.796 & 11.3 &   \nodata &  6.0 &  \nodata \\
AT Ser         &   $-$2.05 &  0.746568 & 5326.814 & 11.0 & \nodata & 12.1 &       1.5 &  6.0 &      1.0 \\
W Tuc          &   $-$1.64 &  0.642243 & 5454.570 & 10.9 &   3.810 & 11.2 &   \nodata &  6.0 &  \nodata \\
\multicolumn{10}{c}{MP RRab Blazhko\tablenotemark{c}} \\
BS Aps         &   $-$1.51 &  0.582560 & 3836.378 & 11.9 &   3.808 & 11.0 &       0.7 &  5.8 &      0.4 \\
DN Aqr         &   $-$1.77 &  0.633766 & 5364.300 & 10.8 &   3.785 & 11.5 &       1.6 &  6.1 &      1.1 \\
S Ara\tablenotemark{b}          
               &   \nodata &  0.451883 & 2137.338 & 10.3 &   3.827 &  8.2 &       0.8 &  3.7 &      0.6 \\
RV Cap         &   $-$1.72 &  0.447748 & 5328.400 & 10.6 &   3.825 & 10.6 &   \nodata &  5.6 &  \nodata \\
RV Cet         &   $-$1.32 &  0.623421 & 5455.460 & 10.6 &   3.804 & 11.2 &       0.6 &  6.0 &      0.4 \\
UV Oct         &   $-$1.69 &  0.542573 & 3836.795 &  9.2 &   3.815 &  9.9 &       1.3 &  5.0 &      0.9 \\
V Ind          &   $-$1.62 &  0.479602 & 1873.466 &  9.6 &   3.825 &  8.6 &       0.4 &  4.1 &      0.3 \\
V1645 Sgr      &   $-$1.81 &  0.552910 & 3932.480 & 11.0 &   3.810 & 11.1 &       0.6 &  5.9 &      0.4 \\
CD Vel         &   $-$1.71 &  0.573509 & 3835.914 & 11.7 &   3.811 & 10.7 &       0.4 &  5.6 &      0.3 \\
AM Vir         &   $-$1.45 &  0.615085 & 1888.395 & 11.2 &   3.806 & 10.8 &       0.8 &  5.7 &      0.5 \\
AS Vir         &   $-$1.59 &  0.553400 & 3836.325 & 11.7 &   3.812 & 10.5 &       0.8 &  5.5 &      0.5 \\
\multicolumn{10}{c}{MR RRab Stable} \\
W Crt          &   $-$0.76 &  0.412013 & 1871.640 & 10.9 &   3.836 &  9.6 &   \nodata &  4.8 &  \nodata \\
DX Del         &   $-$0.52 &  0.472619 & 5323.417 &  9.8 &   3.822 & 11.2 &       0.6 &  5.9 &      0.4 \\
V445 Oph       &   $-$0.05 &  0.397024 & 1939.030 & 10.5 &   3.831 &  9.3 &       0.8 &  4.6 &      0.6 \\
AV Peg         &   $-$0.14 &  0.390382 & 5360.677 & 10.0 &   3.836 &  8.8 &       1.0 &  4.1 &      0.7 \\
HH Pup         &   $-$0.95 &  0.390745 & 1869.661 & 10.6 &   3.837 &  8.5 &       0.5 &  4.0 &      0.1 \\
AN Ser         &   $-$0.06 &  0.522072 & 2701.153 & 10.5 &   3.811 &  7.6 &   \nodata &  3.3 &  \nodata \\
ST Vir         &   $-$0.85 &  0.410806 & 5323.225 & 11.0 &   3.806 & 10.0 &   \nodata &  5.1 &  \nodata \\
UU Vir         &   $-$0.93 &  0.475609 & 1886.488 & 10.1 &   3.826 &  9.8 &       0.2 &  5.0 &      0.1 \\
\multicolumn{10}{c}{MP RRc Stable} \\
as014500-3003  &   $-$2.28 &  0.377380 & 1869.252 & 11.1 &   3.851 & 13.1 &      0.10 &  7.4 &     0.06 \\
as023706-4257  &   $-$1.88 &  0.311326 & 1869.352 &  8.8 &   3.851 & 11.4 &      0.38 &  6.1 &     0.26 \\
as094541-0644  &   $-$2.10 &  0.350225 & 1869.729 & 10.6 &   3.851 & 17.1 &      0.19 &  9.8 &     0.12 \\
as095328+0203  &   $-$1.76 &  0.324697 & 1869.701 &  9.9 &   3.851 & 11.4 &      0.03 &  6.1 &     0.02 \\
as101332-0702  &   $-$1.73 &  0.316000 & 1869.668 & 10.8 &   3.851 & 15.8 &      0.77 &  9.2 &     0.48 \\
as123811-1500  &   $-$1.39 &  0.329045 & 1884.454 & 11.4 &   3.851 & 13.2 &      0.18 &  7.3 &     0.12 \\
as132225-2042  &   $-$0.96 &  0.235934 & 1888.578 & 10.7 &   3.851 & 12.8 &   \nodata &  7.0 &  \nodata \\
as132448-0658  &   $-$2.04 &  0.343237 & 1900.300 & 11.4 &   3.851 & 16.3 &      0.12 &  9.3 &     0.07 \\
as143322-0418  &   $-$1.48 &  0.249632 & 1907.348 & 10.6 &   3.851 &  4.8 &      0.25 &  1.3 &     0.32 \\
as190212-4639  &   $-$2.58 &  0.316900 & 1954.795 &  8.8 &   3.851 &  7.7 &      0.07 &  3.4 &     0.06 \\
as203145-2158  &   $-$1.17 &  0.310712 & 1873.230 & 11.3 &   3.851 & 12.5 &      0.02 &  6.8 &     0.02 \\
as211932-1507  &   $-$1.50 &  0.273458 & 1873.307 & 11.1 &   3.851 &  9.9 &   \nodata &  5.0 &  \nodata \\
\multicolumn{10}{c}{MP RRc Blazhko\tablenotemark{d}} \\
as110522-2641  &   $-$1.60 &  0.294457 & 4900.170 & 11.7 &   3.851 & 15.6 &   \nodata &  8.8 &  \nodata \\
as162158+0244  &   $-$1.83 &  0.323698 & 4900.435 & 12.6 &   3.851 & 25.1 &   \nodata & 12.1 &  \nodata \\
as200431-5352  &   $-$2.66 &  0.300241 & 4900.470 & 11.0 &   3.851 & 15.4 &      2.00 &  8.7 &     1.28 \\
\multicolumn{10}{c}{MP RHB} \\
CS22186-005    &   $-$2.72 & \nodata   & \nodata  & 13.0 &   3.796 & 15.14 &  \nodata & 8.44 &  \nodata \\
CS22190-007    &   $-$2.57 & \nodata   & \nodata  & 14.2 &   3.756 & 11.73 &  \nodata & 6.64 &  \nodata \\
CS22191-029    &   $-$2.70 & \nodata   & \nodata  & 14.1 &   3.778 & 14.99 &  \nodata & 8.42 &  \nodata \\
CS22875-029    &   $-$2.65 & \nodata   & \nodata  & 13.7 &   3.778 & 15.31 &  \nodata & 8.57 &  \nodata \\
CS22878-121    &   $-$2.37 & \nodata   & \nodata  & 14.0 &   3.736 & 10.32 &  \nodata & 5.71 &  \nodata \\
CS22879-097    &   $-$2.45 & \nodata   & \nodata  & 14.2 &   3.763 & 13.16 &  \nodata & 7.46 &  \nodata \\
CS22879-103    &   $-$2.10 & \nodata   & \nodata  & 14.3 &   3.762 & 14.80 &  \nodata & 8.27 &  \nodata \\
CS22881-039    &   $-$2.68 & \nodata   & \nodata  & 15.1 &   3.785 & 13.62 &  \nodata & 7.59 &  \nodata \\
CS22882-001    &   $-$2.48 & \nodata   & \nodata  & 14.8 &   3.775 & 15.21 &  \nodata & 8.45 &  \nodata \\
CS22883-037    &   $-$1.89 & \nodata   & \nodata  & 14.7 &   3.771 & 15.72 &  \nodata & 8.93 &  \nodata \\
CS22886-043    &   $-$2.15 & \nodata   & \nodata  & 14.7 &   3.778 & 15.24 &  \nodata & 8.47 &  \nodata \\
CS22888-047    &   $-$2.34 & \nodata   & \nodata  & 14.6 &   3.778 & 14.93 &  \nodata & 8.34 &  \nodata \\
CS22891-184    &   $-$2.55 & \nodata   & \nodata  & 13.8 &   3.748 & 12.20 &  \nodata & 6.90 &  \nodata \\
CS22896-110    &   $-$2.66 & \nodata   & \nodata  & 13.6 &   3.740 & 10.88 &  \nodata & 6.02 &  \nodata \\
CS22898-043    &   $-$2.84 & \nodata   & \nodata  & 14.1 &   3.782 & 15.31 &  \nodata & 8.59 &  \nodata \\
CS22937-072    &   $-$2.73 & \nodata   & \nodata  & 14.0 &   3.736 & 10.13 &  \nodata & 5.62 &  \nodata \\
CS22940-077    &   $-$2.92 & \nodata   & \nodata  & 14.1 &   3.732 & 10.65 &  \nodata & 5.95 &  \nodata \\
CS22940-121    &   $-$2.89 & \nodata   & \nodata  & 14.2 &   3.732 & 10.34 &  \nodata & 5.69 &  \nodata \\
CS22945-056    &   $-$2.90 & \nodata   & \nodata  & 14.1 &   3.763 & 14.93 &  \nodata & 8.40 &  \nodata \\
CS22948-006    &   $-$2.62 & \nodata   & \nodata  & 15.1 &   3.744 & 11.60 &  \nodata & 6.51 &  \nodata \\
CS22951-077    &   $-$2.42 & \nodata   & \nodata  & 13.6 &   3.728 & 10.27 &  \nodata & 5.66 &  \nodata \\
CS22955-174    &   $-$2.88 & \nodata   & \nodata  & 14.4 &   3.752 & 14.15 &  \nodata & 8.08 &  \nodata \\
\enddata                                                            

\tablenotetext{a}{For RRab stars the log \teff\ values are means over the 
                  pulsational cycles, as given by \cite{skarka14}.
                  For RRc stars we have adopted a constant for log \teff.}
\tablenotetext{b}{We gathered du Pont echelle spectra for this star but did 
                  not include it in our model 
                  atmosphere analyses in previous papers.}

\tablenotetext{c}{Nominal periods; see Table~\ref{tab3} and text.}

\tablenotetext{d}{Blazhko variation reported by \cite{szczygiel07}.}
                                                                    
\end{deluxetable}                                                   
\end{center}

\clearpage
\begin{center}
\begin{deluxetable}{ccr}
\tabletypesize{\footnotesize}
\tablewidth{0pt}
\tablecaption{Atomic Lines\tablenotemark{a}\label{tab2}}
\tablecolumns{3}
\tablehead{
\colhead{$\lambda$}                      &
\colhead{Species}                        &
\colhead{$EW_\odot$\tablenotemark{b}}    \\
\colhead{\AA}                            &
\colhead{}                               &
\colhead{m\AA}                            
}
\startdata
  4045.82 & \species{Fe}{i}  &  1174 \\
  4062.45 & \species{Fe}{i}  &    98 \\
  4063.60 & \species{Fe}{i}  &   787 \\
  4067.99 & \species{Fe}{i}  &   133 \\
  4071.75 & \species{Fe}{i}  &   723 \\
  4076.64 & \species{Fe}{i}  &   127 \\
  4077.72 & \species{Sr}{ii} &   428 \\
  4121.32 & \species{Co}{i}  &   125 \\
  4134.68 & \species{Fe}{i}  &   129 \\
  4143.42 & \species{Fe}{i}  &   134 \\
\enddata

\tablenotetext{a}{This table is available in its entirety in 
                  in the on-line article}

\tablenotetext{b}{Equivalent widths from the \cite{moore66} Solar Atlas}

\end{deluxetable}
\end{center}

\clearpage
\begin{center}
\begin{deluxetable}{lccr}
\tablewidth{0pt}
\tablecaption{Ephemerides for Blazhko Stars Used in Figure~\ref{fig10}\label{tab3}}
\tablecolumns{4}
\tablehead{
\colhead{Star}                           &
\colhead{HJD$_0$}                        &
\colhead{$P$}                            &
\colhead{$\delta\phi$}                    
}
\startdata
AM Vir    & 1888.395 & 0.615085 &   $-$0.010 \\
AS Vir    & 3836.325 & 0.553400 &      0.000 \\
BS Aps    & 3836.370 & 0.582560 &      0.015 \\
RV Cap    & 5328.400 & 0.447748 &   $-$0.010 \\
RV Cet    & 5455.460 & 0.623421 &      0.025 \\
S Ara     & 2137.338 & 0.451883 &   $-$0.030 \\
UV Oct    & 3836.795 & 0.542573 &      0.010 \\
V1645 Sgr & 3932.480 & 0.552910 &   $-$0.010 \\
V Ind     & 1873.466 & 0.479602 &   $-$0.030 \\
V494 Sco  & 5364.435 & 0.427266 &   $-$0.045 \\
\enddata

\end{deluxetable}
\end{center}

\clearpage
\begin{center}
\begin{deluxetable}{cc}
\tablewidth{0pt}
\tablecaption{Mean Values of \vmacrot/\vmic\ for MP RRab Stars\label{tab4}}
\tablecolumns{2}
\tablehead{
\colhead{$\phi$}                         &
\colhead{$\langle$\vmacrot/\vmic$\rangle$}
}
\startdata
0.025  &  4.37 \\
0.075  &  3.71 \\
0.125  &  3.11 \\
0.175  &  2.93 \\
0.225  &  2.55 \\
0.275  &  2.27 \\
0.325  &  1.91 \\
0.375  &  1.81 \\
0.425  &  2.08 \\
0.475  &  2.40 \\
0.525  &  2.69 \\
0.575  &  2.98 \\
0.625  &  3.28 \\
0.675  &  3.41 \\
0.725  &  3.23 \\
0.775  &  3.31 \\
0.825  &  3.41 \\
0.875  &  3.78 \\
0.925  &  4.05 \\
0.975  &  4.32 \\
1.025  &  4.37 \\
\enddata                                                      
                                                              
\end{deluxetable}                                             
\end{center}

\clearpage
\begin{center}
\begin{deluxetable}{lcccc}
\tablewidth{0pt}
\tablecaption{\teff\ and \vrot\ Values for RHB Stars in Two Previous Studies\label{tab5}}
\tablecolumns{5}
\tablehead{
\colhead{Source}                         &
\colhead{Star}                           &
\colhead{\teff}                          &
\colhead{log \teff}                      &
\colhead{\vrot}
} 
\startdata
\mbox{\cite{behr03b}}   &  HD 25532  &  5553 &  3.745 &   7.7 \\
\mbox{\cite{carney08}}  &  HD 25532  &  5320 &  3.726 &   4.8 \\
\mbox{Difference}       &            &   233 &  0.019 &   2.9 \\
                        &            &       &        &       \\
\mbox{\cite{behr03b}}   &  HD 184266 &  5760 &  3.760 &   9.3 \\ 
\mbox{\cite{carney08}}  &  HD 184266 &  5490 &  3.740 &   5.0 \\
\mbox{Difference}       &            &   233 &  0.021 &   4.3 \\
                        &            &       &        &       \\
\mbox{Mean Difference}  &            &   252 &  0.020 &   3.6 \\
\enddata

\end{deluxetable}
\end{center}

\clearpage
\begin{center}
\begin{deluxetable}{lccc}
\tablewidth{0pt}
\tablecaption{Metallicity Data for Three RHB Samples\label{tab6}}
\tablecolumns{4}
\tablehead{
\colhead{Source}                         &
\colhead{$\langle$[Fe/H]$\rangle$}       &
\colhead{$\sigma$}                       &
\colhead{$f$([Fe/H]$>-$1.7)}                
} 
\startdata
\mbox{This study}       &    $-$2.59 &   0.27 &   0.00 \\
\mbox{\cite{carney08}}  &    $-$1.95 &   0.35 &   0.17 \\
\mbox{\cite{behr03b}}   &    $-$1.53 &   0.36 &   0.80 \\
\enddata

\end{deluxetable}
\end{center}

\end{document}